\documentclass[12pt]{article} 
\usepackage{graphicx,subfigure,latexsym, color}
\usepackage{amsmath,amssymb,amsfonts}
\usepackage{bm}

\newcommand{\be}{\begin{equation}}
\newcommand{\ee}{\end{equation}}
\newcommand{\bear}{\begin{eqnarray}}
\newcommand{\eear}{\end{eqnarray}}
\newcommand{\ba}{\begin{array}}
\newcommand{\ea}{\end{array}}

\def \be {\begin{equation}}
\def \ee {\end{equation}}

\def \bes {\begin{subequations}}
\def \ees {\end{subequations}}

\def \<{\langle}
\def \>{\rangle}
\def \+{\dagger}
\def \({\left(}
\def \){\right)}
\def \[{\left[}
\def \]{\right]}

\usepackage[normalem]{ulem}  
\newcommand{\changed}[1]{{\sf\color[rgb]{1,0,0}{#1}}}

\renewcommand\sout{\bgroup \color{red} \ULdepth=-.5ex \ULset}

\usepackage[colorlinks=true, pdfstartview=FitV, linkcolor=red, citecolor=blue, urlcolor=blue]{hyperref}

\begin{document}

\begin{titlepage}
\vfill
\begin{flushright}
{\normalsize RBRC-1205, INT-PUB-16-034}\\
\end{flushright}

\vfill
\begin{center}
{\Large\bf Longitudinal Conductivity in Strong Magnetic Field in Perturbative QCD:  Complete Leading Order}

\vskip 0.3in

\vskip 0.3in
Koichi Hattori$^{1,2}$\footnote{e-mail: {\tt koichi.hattori@riken.jp}},
Shiyong Li$^{3}$\footnote{e-mail: {\tt sli72@uic.edu}},
Daisuke Satow$^{4}$\footnote{e-mail: {\tt dsato@th.physik.uni-frankfurt.de}},
Ho-Ung Yee$^{3,2}$\footnote{e-mail: {\tt hyee@uic.edu}}

\vskip 0.15in

 {\it $^{1}$ Physics Department, Fudan University, Shanghai 200433, China}\\[0.15in]
 {\it $^{2}$ RIKEN-BNL Research Center, Brookhaven National Laboratory, Upton,
 New York 11973-5000, U.S.A.}\\[0.15in]
{\it $^{3}$ Department of Physics, University of Illinois, Chicago, Illinois
 60607, U.S.A.}\\[0.15in]
 {\it $^{4}$ Institut f\"ur Theoretische Physik, Johann Wolfgang Goethe-Universit\"at,
Max-von-Laue-Str\changed{asse} 1, D-60438 Frankfurt am Main, Germany}\\[0.15in]

\end{center}

\vfill

\begin{abstract}

We compute the longitudinal electrical conductivity in the presence of strong background magnetic field in complete leading order of perturbative QCD, based on the assumed hierarchy of scales $\alpha_s eB\ll (m_q^2,T^2)\ll eB$.
We formulate an effective kinetic theory of lowest Landau level quarks with the leading order QCD collision term arising from
1-to-2 processes that become possible due to 1+1 dimensional Landau level kinematics. In small $m_q/T\ll 1$ regime,
the longitudinal conductivity behaves as $\sigma_{zz}\sim e^2(eB)T/(\alpha_s m_q^2\log(T/m_q))$, where the quark mass dependence can be understood from the chiral anomaly with the axial charge relaxation provided by a finite quark mass $m_q$. We also present parametric estimates for the longitudinal and transverse ``color conductivities'' in the presence of strong magnetic field, by
computing dominant damping rates for quarks and gluons that are responsible for color charge transportation.
We observe that the longitudinal color conductivity is enhanced by strong magnetic field, which implies that the sphaleron transition rate in perturbative QCD is suppressed by strong magnetic field due to the enhanced Lenz's law in color field dynamics.

\end{abstract}

\vfill

\end{titlepage}
\setcounter{footnote}{0}

\baselineskip 18pt \pagebreak
\renewcommand{\thepage}{\arabic{page}}
\pagebreak

\section{Introduction and Summary}

In this work we compute the longitudinal electric conductivity of a deconfined QCD quark-gluon plasma in the presence of
a strong background magnetic field in complete leading order of perturbative QCD. 
The motivation comes from either the ultra-relativistic heavy-ion collisions where a strong, albeit short-lived, magnetic field of strength $eB\sim (200 \,{\rm MeV})^2$ is created on top of a deconfined quark-gluon plasma fireball \cite{Kharzeev:2007jp, Bestimates} 
(see Ref.~\cite{Hattori:2016emy} for recent reviews), 
or from the possible quark matter phase in the neutron star core 
with temperature much smaller than the magnetic field $T^2\ll eB$ (see Ref.~\cite{Lai:2000at} for a review). 
Another system where our study may find its relevance is a condensed matter system of Dirac semimetal in a magnetic field\footnote{In Ref.~\cite{Li:2014bha}, a magnetic field with the magnitude $B=2\sim 9$ T (corresponding to the energy scale of $\sqrt{eB\hbar c^2}=11\sim 23$ eV) was introduced to the Dirac semimetal at $k_B T=1.7$ meV.}.
The QCD phase diagram and the phase transitions in such strong magnetic fields have been investigated 
(see Refs.~\cite{Bali:2011qj,Bali:2013esa,Fukushima:2012kc,Braun:2014fua, Hattori:2015aki} and references therein). 
Although the latter system is a dense matter with high quark chemical potential, we will consider a neutral quark-gluon plasma in this work that is relevant more in the heavy-ion collisions.
See a recent work of Ref.\cite{Harutyunyan:2016rxm} that studies the magnetized dense matter case of the neutron star physics, considering QED interactions between electrons and dense nuclear matter,  Ref.\cite{Mamo:2013efa} for a computation in strong coupling limit via AdS/CFT correspondence, and Ref.\cite{Buividovich:2010tn,Buividovich:2010qe} for the lattice QCD computation both in vacuum and at finite temperature for realistic values of magnetic field in heavy-ion collisions.

We will assume a hierarchy of scales that is consistent with perturbative QCD coupling expansion: $\alpha_s eB\ll T^2\ll eB$.
This assumption, introduced in Ref.\cite{Fukushima:2015wck}, leads to a consistent Hard Thermal Loop (HTL) power counting scheme.
The second inequality, which is what we mean by strong magnetic field, allows us to focus on only the lowest Landau level states (LLL) of quarks and antiquarks, since higher Landau level thermal occupation is exponentially suppressed by $e^{-{\sqrt{eB}\over T}}$ and do not participate the transport phenomena in leading order.
The first inequality is more of a theoretical assumption: the dominant charge carriers that contribute to the transport coefficients in leading order are ``hard'' quasi particles of typical momenta $\sim T$, and their dispersion relation deviates
from the free one by $\alpha_s eB/p^2\sim \alpha_s eB/T^2$, since the leading thermal self energy goes as $\Sigma\sim \alpha_s eB$ due to the other inequality $T^2\ll eB$ (the dominant contribution to the 1-loop self energy comes from the LLL states due to their larger density of states than the gluons). The first inequality allows us to neglect these corrections for hard particles in leading order.
The outcome is a consistent HTL scheme with thermally excited ``hard'' LLL states as the dominant source of HTL self-energies
(the density of states for LLL is $\sim (eB)T$ while that for gluons is only $T^3$).

We will introduce a finite quark mass to have a finite longitudinal conductivity with a background magnetic field. 
In the massless limit, the axial anomaly tells us that the axial charge should increase when we apply a longitudinal electric
field as
\be
\partial_t n_A= {e^2N_c N_F\over 2\pi^2} {\bm E}\cdot{\bm B}\,,\label{axial}
\ee
and the Chiral Magnetic Effect \cite{Fukushima:2008xe} current from this axial charge grows linearly in time
\be
{\bm J}={e^2N_cN_F\over 2\pi^2}\mu_A {\bm B}={e^2 N_c N_F \over 2\pi^2\chi}n_A{\bm B}={e^4 (N_c N_F)^2 B^2\over 4\pi^4 \chi}t E\,,
\ee
where $\chi$ is the charge susceptibility.
To have a finite conductivity, we should consider relaxation dynamics of the axial charge: either sphaleron transitions or a finite quark mass \cite{Grabowska:2014efa}. With the relaxation term of $-{1\over \tau_R}n_A$ in the right-hand side of (\ref{axial}), we have a stationary solution $n_A= {e^2N_c N_F\over 2\pi^2} {\bm E}\cdot{\bm B}\tau_R$, which gives a finite contribution to the longitudinal conductivity from Chiral Magnetic Effect \cite{Son:2012bg},
\be
\sigma_{zz}={e^4 (N_c N_F)^2 B^2\over 4\pi^4 \chi}\tau_R\,.\label{sLint}
\ee

The inverse relaxation time from sphaleron dynamics is related to the sphaleron transition rate $\Gamma_s$ by a fluctuation-dissipation relation \cite{Giudice:1993bb}
\be
{1\over \tau_{R,s}}={(2N_F)^2\Gamma_s\over 2\chi T}\,,
\ee
and the sphaleron transition rate without magnetic field is known to be of order $\Gamma_s\sim \alpha_s^5\log(1/\alpha_s)T^4$ \cite{Bodeker:1998hm,Arnold:1998cy,Moore:2010jd}. We will discuss in section \ref{sec5} a possible modification of $\Gamma_s$ in the strong magnetic field, but let us mention here that the result is a further {\it suppression} of $\Gamma_s$, mainly due to an enhanced Lenz's law from the increased color conductivity along the magnetic field direction (while transverse color conductivity remains as $\sigma_{c}\sim{T}$ (neglecting any logarithms in power counting)). This weak coupling behavior is different from the strong coupling one from AdS/CFT correspondence \cite{Basar:2012gh}.

On the other hand, the inverse relaxation time from a finite quark mass goes as
\be
{1\over \tau_{R,m}}\sim \alpha_s m_q^2 /T\,,
\ee
either without or with the strong magnetic field. In the case without magnetic field, it can be shown that the dominant chirality flipping transition rate comes from the small angle scatterings with soft transverse space like magnetic degrees of freedom \cite{misha-yee}, that is, the same one for the leading damping rate of hard particles. This results in a single power of $\alpha_s$ rather than $\alpha_s^2$. The $m_q^2$ dependence is easy to understand since chirality flipping amplitude should be proportional to the mass. On the other hand, in the case with strong magnetic field in our LLL approximation for quarks, since the LLL states have 1+1 dimensional dispersion relation, it becomes possible for an on-shell gluon to pair create quark/antiquark pair and vice versa \cite{Tuchin:2010gx, Fukushima:2011nu, Hattori:2012je}.
This 1-to-2 (and 2-to-1) process rate is only of $\alpha_s$, and becomes dominant over the usual 2-to-2 processes,
(under the assumption $\alpha_s eB\ll m_q^2$ that we will explain later). The resulting chirality flipping rate is again expected to be $\alpha_s m_q^2/T$.
In fact, this is what we compute in this work, confirming this expectation by an explicit computation\footnote{
Two of us (K.H. and D.S.) also evaluate the conductivity in a complementary paper~\cite{Hattori-Satow}, 
by using diagrammatic method instead of the kinetic approach.} (see our section \ref{sec4}).

Because the largest inverse relaxation time determines the final inverse relaxation time, 
a finite quark mass will be dominant over the sphaleron dynamics if $\alpha_s m_q^2\gg \alpha_s^5 T^2$: a condition that can be justified with a small enough coupling.
We will assume this to neglect non-perturbative sphaleron dynamics, focusing only on perturbative quasi-particle dynamics
of LLL quarks interacting with 3+1 dimensional thermal gluons. With $\tau_R\sim T/(\alpha_s m_q^2 )$ in (\ref{sLint}),
and recalling that the charge susceptibility of LLL states in strong magnetic field limit is given by
\be
\chi=N_c {1\over 2\pi}\left(eB\over 2\pi\right)\,,
\ee
where the first $1/(2\pi)$ is the 1+1 dimensional charge susceptibility, and $(eB/2\pi)$ is the transverse density of states of the LLL, we expect to have the longitudinal electric conductivity in small quark mass limit $m_q\to 0$ as
\be
\sigma_{zz}\sim {e^2 N_c}{(eB)  T}{1\over \alpha_s m_q^2}\,,\quad m_q\to 0\,.
\ee
Our computation with the explicit result (\ref{smallm}) indeed confirms this expectation, up to a logarithmic correction of $1/\log(T/m_q)$. 

We will provide a full result of $\sigma_{zz}$ for an arbitrary value of ${m_q/T}$ in complete leading order in $\alpha_s$, under the assumed hierarchy $\alpha_s eB\ll (T^2, m_q^2)\ll eB$. The result takes a form
\be
\sigma_{zz}=e^2{{\rm dim} R\over  C_2(R)}\left(eB\over 2\pi\right){1\over \alpha_s T}\sigma_L(m_q/T)\,,
\ee
with a dimensionless function $\sigma_L(m_q/T)$ given by (\ref{sL}).

In the other case of $m_q^2\ll \alpha_s eB$, which is a quite interesting problem for future, the situation is
complicated since some non-chirality flipping 2-to-2 processes become of the same order as the above 1-to-2 processes\footnote{{{Nevertheless, at the leading-log approximation, the 2-to-2 process is negligible compared to the 1-to-2 process even when $m_q^2\ll \alpha_s eB$.
This case is analyzed in the complementary paper~\cite{Hattori-Satow} at the leading-log accuracy.
}}} (see Appendix 2). 
The chirality flipping processes is still the major ingredient for the final conductivity
(otherwise the conductivity diverges as seen in the above): essentially, these chirality flipping processes
are the ``bottle-neck'' for the relaxation of axial charges that would grow with anomaly, and should be included in the kinetic theory. 

The real complication arises when $m_q\ll\alpha_s T$: in the small momentum region $p_z\sim m_q$, the chirality is maximally violated
and chirality can effectively be flipped by going through this IR region. In 1+1 dimensions, the phase space for this IR region
gives only one power of $m_q$: $\int^{m_q} p_z\sim m_q$, which means that the effective chirality flipping rate from this IR region is suppressed only by a single power $m_q$, $1/\tau\sim \alpha_s^2 m_q$, thwarting the above $\alpha_s m_q^2/T$ chirality flipping rates from hard momentum region when $m_q\ll \alpha_s T$. What this all means is that in 1+1 dimensions with $m_q\ll\alpha_s T$, the major ``bottle-neck'' for axial charge
relaxation happens in the IR region near the origin $p_z\sim m_q$, and this IR dynamics determines the global shape of the distribution function and the final conductivity. Topologically, the two large $p_z$ regions, $p_z>0$ and $p_z<0$,
are connected by the IR region of $p_z=0$, and without knowing the boundary condition at $p_z=0$, one cannot determine
the global solution uniquely.
Since the self energy is of order $\alpha_s eB$, the dispersion relation for these IR modes of $p\sim m_q$ gets thermal correction of $\alpha_s eB/m_q^2\gg 1$, and we no longer should use kinetic theory with free dispersion relation for these IR modes. 
We leave this problem to a future study.

\section{Effective kinetic theory in the LLL approximation}

In this section, we set up our computational framework based on weakly interacting quasi-particles described by kinetic theory. 
We consider a one-flavor case until Sec.~\ref{sec4}, as an extension to the multiflavor case is straightforward. 
In the strong background magnetic field $eB\gg T^2$, the effect of magnetic field on the motions of quarks and antiquarks should be taken care of non-perturbatively, and we achieve this by quantizing the quark field in the presence of background magnetic field, which is summarized in the Appendix 1. The quark wave-functions are now the Landau levels
whose density of states in the transverse two dimensions perpendicular to the magnetic field is $eB/(2\pi)$.
In the Landau gauge $A^2=B x^1$, there are two momentum quantum numbers for each Landau level states: the momentum $p_2$ along $x^2$ direction, and the longitudinal momentum $p_z$ along the direction of magnetic field.
The $p_2$ serves as a label for the transverse position of each Landau levels, and encodes the transverse density of states $eB/(2\pi)$, while $p_z$ is the conventional momentum for the motion of each state along the longitudinal direction.
Correspondingly, the dispersion relation of quasi-particles is
\be
E_{p_z,n}=\sqrt{p_z^2+2|eB|n+m_q^2}\,,
\ee
where $n=0,1,\cdots$ is the Landau levels and $m_q$ is the bare quark mass.

In weakly coupled regime, a quasi-particle picture should be a good description with regard to how the system responds to 
an external perturbation. 
In our case of strong magnetic field, the fermionic quasi-particles are Landau level quarks and antiquarks that move only along the 1+1 dimensions with the above dispersion relation, while their transverse positions do not change in free limit.
Figuratively, we have a collection of 1+1 dimensional fermion theories distributed in the transverse space with the density $eB/(2\pi)$ for each $n$.
At a finite temperature in equilibrium, each Landau level states are occupied by the usual equilibrium thermal distribution functions. In the regime $eB\gg T^2$, only the lowest Landau levels (LLL) with $n=0$ are populated due to an energy gap of $\Delta\sim \sqrt{eB}\gg T$ for higher Landau levels, and therefore higher Landau levels do not contribute to the transport coefficients in this regime. We will focus only on the LLL in the rest of our paper.

On the other hand, the gluons at leading order are 3+1 dimensional quasi-particles. Their dominant self-energy correction
arising from QCD interactions with thermally populated LLL quarks is of order $\Sigma\sim \alpha_s eB$.
With our assumed hierarchy $\alpha_s eB\ll T^2$, this correction is sub-leading compared to the bare momentum $p\sim T$
for majority of ``hard'' particles of $p\sim T$. Therefore, these hard gluons have the bare dispersion relation at leading order.

As shown in the Appendix 1, the Landau level wave function with $p_2$ is localized around $x_1={p_2/ eB}$ with a width
of order $1/\sqrt{|eB|}$. One can construct a wave packet with a central value of $p_2^{\rm center}$ with a width $\Delta p_2$ that is localized in $x_2^{\rm center}$ (note that there is no velocity associated with $p_2^{\rm center}$ since $\partial E/\partial p_2=0$). Then this wave packet has a spatial width of $\Delta x_2\sim 1/\Delta p_2$. Since $\Delta p_2=\Delta x_1 |eB|$, we
have the transverse uncertainty of $\Delta x_1 \Delta x_2\sim 1/|eB|$, which is the well-known transverse size of the Landau levels.
 An accurate counting of available states in the Appendix 1 shows that the transverse density of such states is $eB/(2\pi)$.
In this way, the label $p_2$ is effectively transformed into a transverse space position variable ${\bm X}_T$ (up to an ambiguity of $1/\sqrt{|eB|}$),
\be
p_2\to (p_2^{\rm center},x_2^{\rm center})\to (x_1^{\rm center},x_2^{\rm center})={\bm X}_T\,.
\ee
This decomposition is similar to the decomposition of space and momentum up to an ambiguity of $\hbar$: the transverse space is roughly a phase space with $\Delta {\bm X}_T^2\sim 1/|eB|$.
We will continue to use $p_2$ for a label for the Landau levels.

Perturbative QCD interactions can induce momentum as well as transverse position changes of each Landau level quasi-particles by scatterings with other quarks/antiquarks or gluons. We will shortly see that in addition to conventional 2-to-2 scatterings that have been considered in literature, we have additional leading 1-to-2 scatterings due to the presence of magnetic field. Explicit computations in section \ref{sec3} and Appendix 2 shows that the changes in $p_2$ due to QCD scatterings is bounded by $\Delta p_2\lesssim \sqrt{eB}$ due to a form factor $R_{00}({\bm q}_\perp)=e^{-{{\bm q}_\perp^2\over 4eB}}$, which means that these interactions are local in the transverse space within a distance of $\Delta p_2/eB\sim 1/\sqrt{eB}$. Therefore if the variation scale of external parameters (such as electric field or temperature gradient)
is much larger than $1/\sqrt{eB}$, we can introduce a further decomposition
\be
p_2= p_2^{\rm global}+\tilde p_2\,,
\ee
where $p_2^{\rm global}$ encodes a large scale (compared to $1/\sqrt{eB}$) transverse position ${\bm X}_T$,
while $\tilde p_2\lesssim \sqrt{eB}$ counts a local collection of Landau levels around ${\bm X}_T$. 
This decomposition is possible, since the theory is invariant under a constant shift of $p_2$.

Based on these, we are led to introduce the 
\changed{quark and anti-quark} distribution functions, 
\be
f_\pm (z, p_z,{\bm X}_T,p_2,n)\,,
\ee
as an occupation number per unit $dzdp_z/(2\pi)$ for the state labeled by $(p_2,n)$ around the global position ${\bm X}_T$.
The $\pm$ refers to quark and antiquark respectively, and for gluons, we have the usual (color diagonal) gluon distribution function $f_g({\bm x},{\bm k})$.
The dynamics of these distributions should be described by the Boltzmann equation,
\be
{\partial f_\pm \over\partial t} + \dot{z}{\partial f_\pm\over\partial z}+\dot{p_z}{\partial f_\pm\over\partial p_z}=C[f_\pm,f_g]\,.
\ee
Note that the dynamical change of $({\bm X}_T,p_2)$ representing the transverse position of Landau level states
should only arise as a result of QCD scatterings, and it is a part of the collision term in the right-hand side.
This reflects the absence of classical transverse motion of Landau level states in free limit: the transverse motions of quark/antiquark are quantum processes.
By the same reason, the above framework is not very useful for computing the transverse conductivities.
Our longitudinal electric conductivity specifically results from the classical longitudinal motion along the momentum $p_z$ induced by an applied electric field:
\be
\dot p_z=\pm eE\,.
\ee
The Feynman rules how to write down the collision terms from the specific QCD interaction diagrams are derived in the Appendix 1, which we will refer to throughout our computations.

In terms of the effective distribution functions $f_\pm$, the electric current from a state of $({\bm X}_T,p_2,n)$ is written as
\be
J_z=e \,{\rm dim} R\int {dp_z\over (2\pi)} \,v_p \,(f_+(p_z,{\bm X}_T,p_2,n)-f_-(p_z,{\bm X}_T, p_2,n))\,,\quad v_p={p_z\over E_p}\,,
\ee
where $v_p$ is the longitudinal velocity and dim R is the dimension of color representation.
This formula is quite intuitive.

For the kinetic theory to be meaningful, the mean free path should be larger than the Compton wavelength of dominant quasi-particles governing the response. For most of transport coefficients, the dominant charge carriers have typical momenta of order temperature $p_z\sim T$ (and the wave-length of $1/T$), while the mean free path
arising from scatterings with other ambient particles are found to be about 
$l_{\rm mfp}\sim T/(\alpha_s m_q^2)$ (for ``kinetic relaxation'' of quarks) 
or $l_{\rm mfp}\sim 1/(\alpha_s T)$ (for ``color'' relaxation). 
These results are derived in section \ref{sec3} and section \ref{sec5}, respectively. 
Therefore, with a sufficiently small $\alpha_s\ll 1$ and a reasonable ratio of $m_q^2/T^2\lesssim {\cal O}(1)$, this criterion is satisfied.

Finally we explain one important aspect in our QCD collision terms.
Due to the 1+1 dimensional dispersion relation for quark/antiquarks and the usual 3+1 dimensional dispersion relation for gluons, it is now kinematically possible to have on-shell 1-to-2 processes of quark/antiquark pair annihilation/pair creation 
to/from a single gluon \cite{Tuchin:2010gx, Fukushima:2011nu, Hattori:2012je}. These are the only on-shell 1-to-2 processes, and their rate is proportional to a single power of $\alpha_s$,
even for the ``kinetic relaxation rate'' relevant for electric conductivity. A more careful computation in section \ref{sec3} shows that it is $\sim \alpha_s m_q^2/T$, featuring a universal factor of $m_q^2$ for both chirality flipping and non-flipping rates. As we show in the Appendix 2, the kinetic relaxation relevant for conductivity arising from the conventional 2-to-2 processes is at most of $\alpha_s^2 (eB/T)\log(m_q^2/\alpha_seB)$\footnote{However, we show in section \ref{sec5} that the damping rate relevant for ``color conductivity'' from 2-to-2 processes is of order $\alpha_s T$, similar to the conventional case.
As the damping rate from 1-to-2 processes is the same $\alpha_s m_q^2/T$, the 2-to-2 processes dominate over 1-to-2 processes for color conductivity when $m_q\lesssim T$.}. With our assumed hierarchy of scales $\alpha_s eB\ll (T^2,m_q^2)\ll eB$, these 2-to-2 processes are sub-leading compared to the novel 1-to-2 processes in the presence of strong magnetic field.

\section{Collision term in leading order\label{sec3}}

In this section we work out the leading order collision term from the pair creation/annihilation processes discussed in the previous section. In the case of longitudinal conductivity, the applied electric field $\bm E$ is parallel to the magnetic field and is homogeneous in the two dimensional transverse space. The resulting distribution functions for 1+1 dimensional Landau levels will be homogeneous in the transverse space too, that is, there will be no dependence on the label of the LLL states, $p_2$ in our Landau gauge:
\be
f_\pm(p_z,p_2)\equiv f_\pm(p_z)\,.
\ee
We will write down the collision term with this assumption for simplicity.

From the Feynman rules summarized in the Appendix 1, it is straightforward to get the collision term for the quark distribution $f_+(p_z)$ as 
\bear
C[f_+(p_z)]&=&{1\over 2 E_p}\int {d^2 {\bm p'}\over (2\pi)^2 2E_{p'}} \int {d^3 {\bm k}\over (2\pi)^3 2E_k} \left|{\cal M}\right|^2
(2\pi)^3\delta^{(2)}\left(p+p'-k) \delta(E_p+E_{p'}-E_k\right)\nonumber\\
&\times&\left(\left(1-f_+(p_z)\right)\left(1-f_-(p_z')\right)f_g(k)-f_+(p_z)f_-(p_z')\left(1+f_g( k)\right)\right)\,,\label{collision}
\eear
where $E_p=\sqrt{p_z^2+m_q^2}$, $E_k=|\bm k|$, $d^2{\bm  p}\equiv dp_z dp_2$, and 
\be
\delta^{(2)}(p)\equiv \delta(p_z)\delta(p_2)\,,
\ee
which includes only {\it spatial} two dimensions $(p_z, p_2)$ (recall that $p_2$ is the label for the LLL states), while
we write down the energy $\delta$-function explicitly. Note that the gluon momentum $\bm k$ is fully three dimensional.
The above is the sum of the pair creation and annihilation processes with the detailed balance condition imposed, such that we can combine them with a common matrix element ${\cal M}$.
The collision term for the antiquark distribution is similar.

Following the conventional treatment, we write down a deviation from the equilibrium in linear order as
\bear
f_\pm(p_z)&=&f^{eq}_F(E_p)+\beta f^{eq}_F(E_p)\left(1- f^{eq}_F(E_p)\right)\chi_\pm(p_z)\,,\nonumber\\
f_g(k)&=&f^{eq}_B(E_k)+\beta f^{eq}_B(E_k)\left(1+ f^{eq}_B(E_k)\right)\chi_g(k)\,,
\eear
with $f_{F/B}^{eq}(\epsilon)=1/(e^{\beta \epsilon}\pm 1)$ and $\beta=1/T$. 
Using the energy $\delta$-function for the detailed balance, we can show that
\bear
&&\left(1-f_+(p_z)\right)\left(1-f_-(p_z')\right)f_g(k)-f_+(p_z)f_-(p_z')\left(1+f_g( k)\right)\nonumber\\&=&\beta f^{eq}_F(E_p)f_F^{eq}(E_{p'})\left(1+f^{eq}_B(E_k)\right) \left(\chi_g(k)-\chi_-(p_z')-\chi_+(p_z)\right)\,.
\eear

In the specific situation we are considering, it is clear that $\chi_-(p_z)=-\chi_+(p_z)$, that is, the effect of the applied electric field on the antiquarks is precisely opposite to the effect on the quarks due to the opposite charge. Similarly, charge conjugation invariance tells us that gluon distribution should not be affected: $\chi_g(k)=0$ \footnote{The magnetic field is C-odd, so breaks C-invariance. 
However, its effects on 1+1 dimensional LLL dynamics depend only on $|eB|$ except the Schwinger phase (see below). 
Since the Schwinger phase is irrelevant for our leading order collision term, we can effectively use C-invariance.}. 
Furthermore, $\bm E$ is a 1 dimensional vector in $\hat z$ space, and $\chi_\pm(p_z)$ is a scalar, which dictates that the response
should take a form
\be
\chi_+(p_z)=(\bm E\cdot p_z)F(|p_z|)=-\chi_+(-p_z)\,,
\ee
that is, $\chi_+(p_z)$ is an odd function on $p_z$. All these facts can be explicitly derived from the structure of the collision term
and the source term from the electric field in the Boltzmann equation. They mirror the similar statements in the 3+1 dimensional case \cite{Arnold:2000dr,Arnold:2003zc}. 

The matrix element from the Feynman rules in the Appendix 1 is given by
\be
{\cal M}=i g_s \epsilon_\mu(k) \left[\bar v(p')\gamma^\mu_\parallel t^a_R u(p)\right]  R_{00}(\bm k_\perp)e^{i\Sigma}\,,
\ee
where $\epsilon(k)$ is the polarization of the external gluon, $t_R$ is the color generator in the quark representation R. 
The phase factor $e^{i\Sigma}$ is called the Schwinger phase \cite{Hattori:2015aki} 
whose expression can be found also in the Appendix, but it will disappear in the matrix square at the end. 
The subscript $\parallel$ denotes 1+1 dimension with the one spatial direction being parallel to $ {\bm B}$,
and 
it is important to note that $\gamma^\mu_\parallel$ is 1+1 dimensional $\gamma$-matrix 
which is effectively $2\times 2$ by the projection operator we are omitting here. Correspondingly, the spinors $u(p)$ and $v(p')$ are 1+1 dimensional spinors with relativistic normalization
\be
u(p)\bar u(p)=\gamma^\mu_\parallel p_\mu+m_q\,,\quad v(p')\bar v(p')=\gamma^\mu_\parallel p'_\mu-m_q\,.
\ee
Finally the form factor originating from the finite transverse size 
$l_B\sim 1/\sqrt{|eB|}$ of the LLL wave function is
\be
R_{00}(\bm k_\perp)=e^{-{\bm k_\perp^2\over 4|eB|}}\,.
\ee
We have to sum $|{\cal M}|^2$ over all incoming antiquark color states and the out-going gluon states, and average over the 
color states of the incoming quark. The color algebra gives a Casimir factor $C_2(R)$ as usual, and the gluon polarization sum 
is
\be
\sum_\epsilon \,\epsilon_\mu (\epsilon_\nu)^*=\delta_{ij}-{\bm k_i \bm k_j\over |\bm k|^2}\,.
\ee
We have
\bear
|{\cal M}|^2&=&g_s^2 C_2(R)e^{-{\bm k_\perp^2\over 2|eB|}}\left(\delta_{ij}-{\bm k_i \bm k_j\over |\bm k|^2}\right)
{\rm Tr}\left[(\gamma^\mu_\parallel p'_\mu-m_q)\gamma^i_\parallel (\gamma^\mu_\parallel p_\mu+m_q)\gamma^j_\parallel\right]\nonumber\\ 
&=& 2g_s^2 C_2(R) e^{-{\bm k_\perp^2\over 2|eB|}}{\bm k_\perp^2\over |\bm k|^2}\left(E_p E_{p'}+p_zp_z'+m_q^2\right)\,.
\label{mat}
\eear
The form factor $e^{-{\bm k_\perp^2\over 2|eB|}}$ reflects the finite transverse size of the LLL states, 
and $\bm k_\perp\ll \sqrt{|eB|}\sim 1/l_B$ modes can not resolve the LLL states.


Without loss of generality, we can choose $p_2=0$ in (\ref{collision}), and perform $p_2'$ and $k_z$ integration to arrive at
\bear
C[f_+(p_z)]&=&{2g_s^2 C_2(R)}{1\over 2E_p}\int {dp_z'\over 2E_{p'}}\int {d^2 {\bm k}_\perp\over (2\pi)^2 2E_k} e^{-{{\bm k}_\perp^2\over 2|eB|}}{\bm k_\perp^2\over |\bm k|^2}\left(E_p E_{p'}+p_zp_z'+m_q^2\right)\nonumber\\
&\times& \delta(E_k-E_p-E_{p'})\beta f^{eq}_F(E_p)f_F^{eq}(E_{p'})\left(1+f^{eq}_B(E_k)\right) \left(\chi_+(p_z')-\chi_+(p_z)\right)\,,
\eear
where it is understood that $k_z=p_z+p_z'$. The energy $\delta$-function can be worked out as
\bear
\delta(E_k-E_p-E_{p'})&=&\delta\left(\sqrt{(p_z+p_z')^2+{\bm k}_\perp^2}-E_p-E_{p'}\right) 
\nonumber
\\
&=& (2 E_k)\,\delta\left({\bm k}_\perp^2-(p_\parallel+p'_\parallel)^2\right)
\eear
and performing ${\bm k}_\perp$ integration, we finally obtain the leading order collision integral
\bear
C[f_+(p_z)]&=&{\alpha_s C_2(R)}m_q^2\int dp_z' {\beta\over E_pE_{p'}}f^{eq}_F(E_p)f_F^{eq}(E_{p'})\left(1+f^{eq}_B(E_p+E_{p'})\right) \left(\chi_+(p_z')-\chi_+(p_z)\right)\,,\nonumber\\
\label{finalcol}
\eear
where we can safely neglect the form factor $e^{-{{\bm k}_\perp^2\over 2|eB|}}$ in our assumed hierarchy $T^2\ll eB$, because ${\bm k}_\perp^2= (p+p')^2\sim T^2$
due to the Boltzmann factor in the collision term which ensures that the dominant leading contribution to the collision integral and the conductivity comes from the hard momentum modes $(p_z,p_z')\sim T$.

It is worthwhile to mention that the above simple collision integrand is a result of the two competing effects.
The first one is the ``chirality selection rule''  that results from the spinor trace in Eq.~(\ref{mat}) 
and gives rise to the factor 
\be
E_p E_{p'}+p_zp_z'+m_q^2 \to |p_z||p_z'|+p_z p_z' 
\label{eq:chirality}
\,,
\ee
where the right-hand side is the expression in the massless limit $(m_q \to 0) $. 
In the massless limit, this factor is nonvanishing only when the $p_z$ and $p_z'$ have the same sign. 
This is related to the chirality selection, because, 
when the spin of the LLL fermions are aligned along the magnetic field, 
the signs of the longitudinal momentum and of the chirality are locked. 
Specifically, the chirality of the LLL massless fermion is identical to the sign of ${\rm sgn} (eB) p_z$:
if we quantize a right-handed spinor field $\psi_R$, both quarks and antiquarks carry only the ${\rm sgn}(eB)p_z>0$ modes.
Likewise, the quarks and antiquarks from a left-handed field $\psi_L$ carry only the ${\rm sgn} (eB)p_z<0$ modes. 
An important fact is that gauge interactions via $\gamma$-matrix do not mix $\psi_R$ and $\psi_L$ fields, which means that
the ${\rm sgn} (eB)p_z>0$ modes and ${\rm sgn} (eB)p_z<0$ modes do not interact in the massless limit. 
Therefore, the chirality selection imposes the longitudinal momenta to have the same sign in Eq.~(\ref{eq:chirality}).

The other selection rule comes from the gluon polarization factor
\be
{{\bm k}_\perp^2\over |\bm k|^2}={(p_\parallel+p'_\parallel)^2\over 
(E_p+E_{p'})^2}=2{(E_p E_{p'}-p_zp_z'+m_q^2)\over (E_p+E_{p'})^2}
\to 
2{(\vert p \vert \vert p'\vert -p_zp_z')\over (\vert p \vert  + \vert p'\vert )^2}
\, ,
\ee
where the last expression is again for the massless limit ($m_q \to 0 $). 
We can understand the role of this factor as follows.
If $\bm k_\perp=0$, that is, the momentum of the external gluon line 
is parallel to the $\hat z$ direction, 
its polarization will be along the transverse space.
However, the motion of the LLL quarks and antiquarks are (1+1)-dimensional and their current is strictly longitudinal, 
and hence the gluon emission vertex should vanish in the $\bm k_\perp=0$ limit. 
In the massless limit, this factor imposes the $p_z$ and $p_z'$ to have the opposite sign. 

We have just found that the fermion-chirality and gluon-polarization selection rules 
impose the competing conditions on the signs of $p_z$ and $p_z'$. 
Actually, they are not compatible in the massless limit, 
so that the collision term vanishes in this limit. 
Indeed, with a finite mass, the product of these two factors, 
and thus the collision term, is proportional to the mass square 
and is independent of the relative sign of $p_z$ and $p_z'$. 
This is an exact form of the mass dependence, 
and shows the universal suppression of the collision integral $C[f_+(p_z)]\sim m_q^2$ 
in the small mass limit.


\section{The longitudinal conductivity in leading order\label{sec4}}

With the leading order collision term in the effective Boltzmann equation in the previous section, 
we are ready to compute the longitudinal electric conductivity in complete leading order.
The Boltzmann equation for the quark distribution in the applied electric field ${\bm E}=E \hat z$ is
\be
{\partial f_+(p_z)\over \partial t}+eE{\partial f_+(p_z)\over\partial p_z}=C[f_+(p_z)]\,.
\ee
We seek a stationary solution in the linear order in $E$ to find the equilibrium conductivity.
From $f_+(p_z)=f^{eq}_F(E_p)+\beta f^{eq}_F(E_p)\left(1- f^{eq}_F(E_p)\right)\chi_+(p_z)$, and 
\be
{\partial E_p\over \partial p_z}\equiv v_p={p_z\over E_p}\,,
\ee
we finally obtain the integral equation for $\chi_+(p_z)$ sourced by the applied electric field,
\bear
&&-eE {p_z\over E_p} f^{eq}_F(E_p)\left(1-f_F^{eq}(E_{p})\right)\nonumber\\
&=&{\alpha_s C_2(R)m_q^2}\int dp_z'{1\over E_pE_{p'}}
 f^{eq}_F(E_p)f_F^{eq}(E_{p'})\left(1+f^{eq}_B(E_p+E_{p'})\right) \left(\chi_+(p_z')-\chi_+(p_z)\right)\,.\label{integral}
\eear
This is a neat one dimensional integral equation which can be solved as follows.
Recall that $\chi_+(p_z')$ is an odd function of $p_z'$, and since the other integrand is an even function of $p_z'$, we see that
the integral with $\chi_+(p_z')$ simply vanishes. Then, $\chi_+(p_z)$ is easily solved as
\be
\chi_+(p_z)={eE \over 2 C_2(R) \alpha_s m_q^2} {p_z\,(1-f_F^{eq}(E_p))\over
\int_0^\infty dp_z' {1\over E_{p'}} f_F^{eq}(E_{p'}) (1+f_B^{eq}(E_p+E_{p'}))}\,.\label{solu}
\ee

In fact, what appears in front of $\chi_+(p_z)$ in (\ref{integral}) is nothing but the quark damping rate
\be
\gamma_q={\alpha_s C_2(R)m_q^2\over E_p \left(1-f_F^{eq}(E_{p})\right)}\int dp_z'{1\over E_{p'}}
f_F^{eq}(E_{p'})\left(1+f^{eq}_B(E_p+E_{p'})\right)\,,\label{qdamp}
 \ee
 which gives a relaxation dynamics in the Boltzmann equation
 \be
 \partial_t\chi_+(p_z)\sim -\gamma_q \,\chi_+(p_z)\,.\label{damp}
 \ee
 Then, the solution (\ref{solu}) is nothing but
 \be
 \chi_+(p_z)={eE}{p_z\over E_p}{1\over\gamma_q}\,,
 \ee
 that is, the relaxation time approximation with the momentum dependent relaxation time $\tau_R=1/\gamma_q$ 
 is in fact an exact solution of the full Boltzmann equation in our special case.

After finding the solution $\chi_+(p_z)$, the longitudinal current $j_z$ is given by
\be
j_z= e \left(eB\over 2\pi\right)2  \,{\rm dim} R\int {dp_z\over (2\pi)} \,v_p \,\beta f^{eq}_F(E_p)\left(1-f_F^{eq}(E_{p})\right)\chi_+(p_z)\,,\quad v_p={p_z\over E_p}\,.
\ee
The factor $(eB/2\pi)$ is the transverse density of LLL states, and the next factor $2$ comes from the equal contribution from the antiquarks. The final expression for our longitudinal conductivity is then 
\be
\sigma_{zz}=  e^2 \left(e B\over 2\pi\right){ {\rm dim} R\over  C_2(R)\alpha_s m_q^2}\int_{-\infty}^{+\infty} {dp_z\over (2\pi)} {p_z^2\over T E_p}
{f_F^{eq}(E_p)(1-f_F^{eq}(E_p))^2\over
\int_0^\infty dp_z' {1\over E_{p'}} f_F^{eq}(E_{p'}) (1+f_B^{eq}(E_p+E_{p'}))}
\,.
\ee

For a general value of $m_q/T$, we need a simple numerical integration to get the result, but the small $m_q$ limit can be
handled more accurately.
In this limit, note that the $p_z'$ integral in the denominator has a logarithmic IR enhancement in $p_z'\sim 0$ regime
due to $1/E_{p'}=1/\sqrt{p_z'^2+m_q^2}$ factor, \changed{as}
\be
\int_0^\infty dp_z' {1\over E_{p'}} f_F^{eq}(E_{p'}) (1+f_B^{eq}(E_p+E_{p'}))\sim {1\over 2} (1+f_B^{eq}(E_p))\log(T/m_q)\quad ({\rm leading\,\,log\,\,in\,\,}m_q/T)\,.
\ee
Using the integral
\be
\int_{-\infty}^\infty {dp_z\over (2\pi)} {p_z^2\over T E_p}{f^{eq}_F(E_p)(1-f^{eq}_F(E_p))^2\over (1+f_B^{eq}(E_p))}={T\over 2\pi}\,,\quad m_q\to 0\,,
\ee
we have the small $m_q$ limit as
\be
\sigma_{zz}\to {e^2\over \pi}{ {\rm dim} R\over  C_2(R)} \left(e B\over 2\pi\right){T\over \alpha_s m_q^2\log(T/m_q)}\,,\quad m_q\to 0\,.\label{smallm}
\ee

\begin{figure}[t]
\centering
\includegraphics[width=0.6\textwidth]{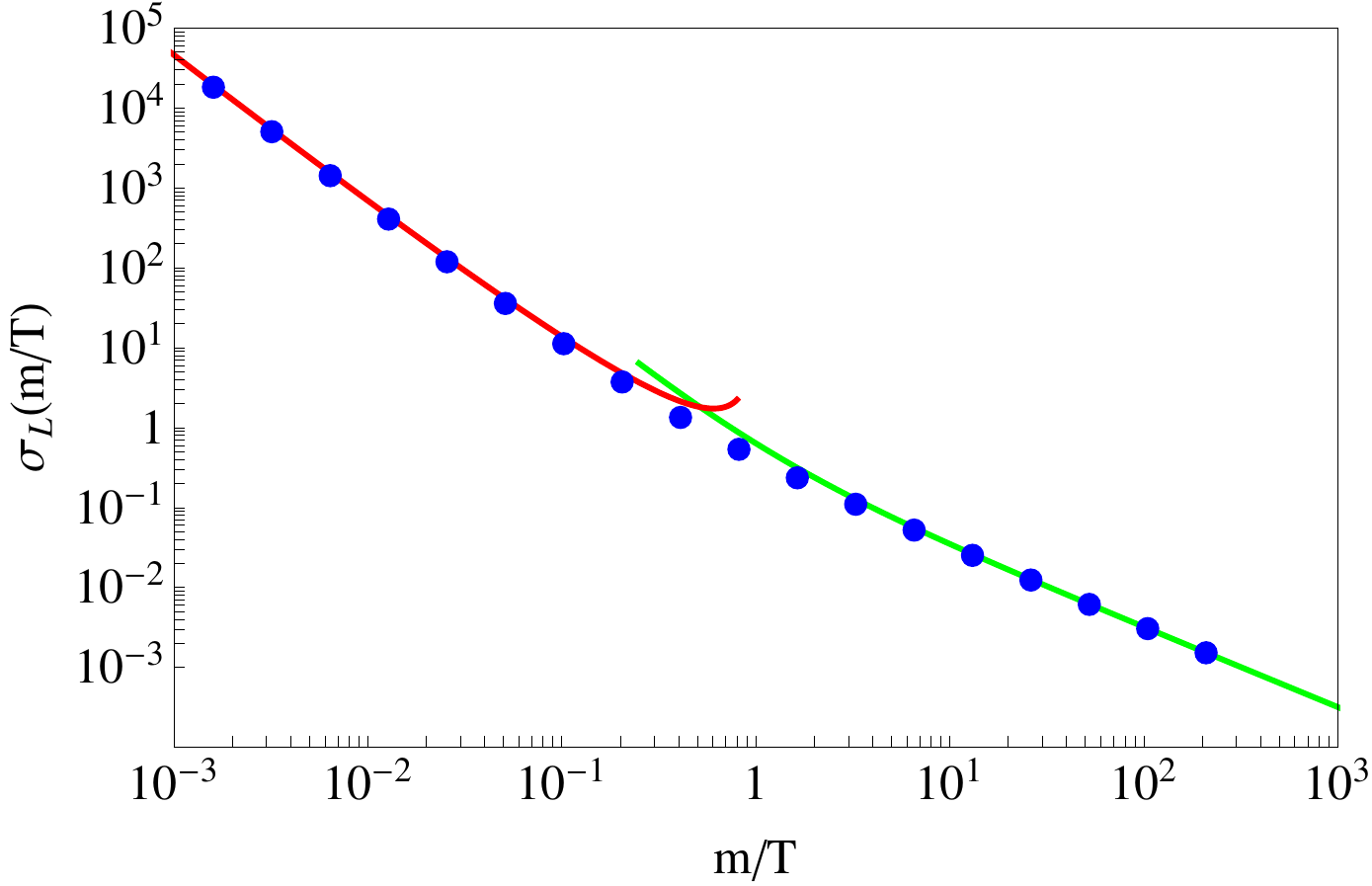}
\caption
{A plot of $\sigma_L(m_q/T)$ from numerical evaluations (blue dots), 
compared to the leading-log expression of $m_q/T$ in Eq.~(\ref{LLLogm}) (red curve) 
and heavy quark limit (\ref{LLLolm}) (green curve).}
\label{fig:sigma}
\end{figure}

Here, we extend our result to the multi-flavor case 
and write the longitudinal conductivity 
in terms of dimensionless variables $\bar p=p_z/T$ and $\bar m =m_q/T$. 
Taking the sum of the flavor dependences arising 
from the electric charge ($ e_f $) and mass ($ \bar m_f$) of the fermion, 
we have 
\be
\sigma_{zz}= \sum_f e_f^2{{\rm dim} R\over  C_2(R)}\left(e_f B\over 2\pi\right){1\over \alpha_s T}\sigma_L(\bar m_f)\,,
\ee
where 
\be
\sigma_L(\bar m)={2\over\bar m^2}\int_0^\infty {d\bar p\over (2\pi)}{\bar p^2\over\epsilon_{\bar p}}{n_F(\epsilon_{\bar p})\left(1-n_F(\epsilon_{\bar p})\right)^2\over \int_0^{\infty} {d\bar p'\over\epsilon_{\bar p'} }n_F(\epsilon_{\bar p'})(1+n_B(\epsilon_{\bar p}+\epsilon_{\bar p'}))}\,,\label{sL}
\ee
and $\epsilon_{\bar p}=\sqrt{\bar p^2+\bar m^2}$ and $n_{F/B}(\epsilon)=1/(e^\epsilon\pm 1)$.
In the small $\bar m\to 0$ limit shown in Eq.~(\ref{smallm}), we have 
\be
\sigma_L(\bar m)\to {1\over \pi \bar m^2 \log(1/\bar m)}\, ,\label{LLLogm}
\ee
while, in the opposite limit ($\bar m\to \infty$), 
\be
\sigma_L(\bar m)\to {1\over \pi \bar m}
\, .
\label{LLLolm}
\ee
Figure~\ref{fig:sigma} shows a plot of $\sigma_L(m_q/T)$ from 
the numerical evaluation, compared to the asymptotic expressions in the two  limits. 
It shows that the leading-log result (\ref{LLLogm}) can be trusted when $m_q/T\lesssim 0.1$, and 
the heavy-quark limit (\ref{LLLolm}) is reliable for $m_q/T \gtrsim 5$.

\section{Damping rates, color conductivity, and the sphaleron rate with strong magnetic field\label{sec5}}

In this section, we study a somewhat different physics of ``color conductivity'' \cite{Selikhov:1993ns} that is an essential ingredient in
computing the sphaleron transition rate in leading order of perturbative QCD \cite{Bodeker:1998hm,Arnold:1998cy}. The color conductivity appears
in the effective Bodeker theory governing ultra-soft color magnetic field dynamics that is responsible for non-perturbative sphaleron transitions. At such low frequency-momentum scales, the color field dynamics reduces to ``magneto-hydrodynamics" where the magnetic fields diffuse at a rate given by the well-known diffusion-type dispersion formula
\be
\omega\sim -i {k^2\over \sigma_c}\,,\label{magdif}
\ee 
where $\sigma_c$ is the color conductivity. The diffusion of magnetic field is resisted by Faraday current, which is also called Lenz's law. Since the Faraday current is proportional to the conductivity $\sigma_c$, the diffusion rate is inversely proportional to the color conductivity $\sigma_c$ in the above. The Bodeker theory is a non-Abelian magneto-hydrodynamics with this color conductivity, with additional thermal noise from the fluctuation-dissipation relation that ensures equilibrium thermal distributions.

The key difference in the physics of color conductivity from the usual abelian conductivity we compute in the previous sections is that even scatterings with small momentum exchange (or small $q_z$ scatterings in the case of LLL quarks) can contribute to
the effective mean-free path of color transportation, since they can change colors without changing the momentum significantly \cite{Selikhov:1993ns}.
This means that the mean-free path for color transportation is determined by the (largest) damping rate, which is roughly a total scattering rate
of a given hard quasi-particle, up to a color charge factor which we will not be precise about.
We will focus only on the parametric dependence of color conductivity on the coupling, magnetic field, temperature, and the quark mass. Denoting the dominant damping rate by $\gamma$, the color conductivity is parametrically given by
\be
\sigma_c\sim \alpha_s \,\,{({\rm density\,of\,states})/T\over\gamma}\,,\label{colorc}
\ee
where $\alpha_s$ in front is a trivial coupling factor in the definition\footnote{More precisely, the numerator is the phase space integral of $-\partial f^{eq}(p)/\partial p=\beta f^{eq}(p)(1\pm f^{eq}(p))$.}.  The majority of this section is devoted to computing the damping rate $\gamma$ for both LLL quarks and the gluons.

In the presence of strong magnetic field, the color conductivity is asymmetric as well. The LLL quarks/antiquarks
can transport the color charges only along the direction of magnetic field, and the only charge carriers in the transverse direction are gluons. On the other hand, the LLL fermions have a larger density of states $(eB)T$ than that for the thermal gluons $T^3$, and moreover we will find that the quark damping rate is parametrically smaller than the gluon damping rate.
Therefore, we conclude that the longitudinal color conductivity is larger than the transverse color conductivity. 
We will discuss the implication of this in the sphaleron transition rate at the end of this section.

The damping rate of a quasi-particle of momentum $p$ can easily be read from the Boltzmann equation for $\chi(p)$ by keeping only $\chi(p)$ term in the collision
term,
dropping all other $\chi$'s with different momenta than $p$. Then the Boltzmann equation gives a relaxation for $\chi(p)$
as
\be
\partial_t \chi(p)=-\gamma_p \chi(p)\,,
\ee
with the damping rate $\gamma_p$ (see (\ref{damp}) as an example). In this way we see how a particular single mode of a momentum $p$ relaxes to the equilibrium with the damping rate.

\subsection*{Quark damping rate}

We will compute the three major contributions to the LLL quark damping rate: 1) 1-to-2 process, 2) 2-to-2 quark-quark/antiquark t-channel scatterings, and 3) 2-to-2 quark-gluon t-channel scatterings. Since t-channel scatterings are expected to be at least larger than s-channel by potential IR enhancement, we think that these computations are enough to identify the parametric dependence of the leading order damping rate of hard quarks.

The collision terms of all these processes are worked out in the other sections: 1) is in section \ref{sec3} and 2),3) are in the Appendix \ref{appendix2}, so we can easily borrow the results from these sections. The damping rate from the 1-to-2 process is
already given in (\ref{damp}), which we reproduce here,
\be\gamma_q^{1-2}={\alpha_s C_2(R)m_q^2\over E_p \left(1-f_F^{eq}(E_{p})\right)}\int dp_z'{1\over E_{p'}}
f_F^{eq}(E_{p'})\left(1+f^{eq}_B(E_p+E_{p'})\right)\sim \alpha_s m_q^2/T\,,\label{qdamp12}
 \ee
where the last expression is our parametric estimate for a hard momentum $E_p\sim T$.

For 2-to-2 quark-quark/antiquark scatterings, we can start from the collision term derived in (\ref{col22}), and the damping rate from this by looking at the coefficient in front of $\chi_+(p_z)$ is written as
\bear
\gamma_q^{q-q}&=&
8\pi\alpha_s^2T_RC_2(R)\left(eB\over 2\pi\right){m_q^4\over E_p} 
\int{dp_z'\over (2\pi)}{1\over E_{p'} |E_pp_z'-E_{p'}p_z|}\nonumber\\&\times&{ 1\over (\changed{{\Lambda}^2_{\text{IR}}}+2(E_pE_{p'}-p_zp_z'-m_q^2))}f^{eq}_F(E_{p'})(1- f^{eq}_F(E_{p'}))\,,\label{qdamp221}
\eear
\changed{where ${\Lambda}^2_{\text{IR}}\sim (\alpha_s eB/m_q^2)^{2\over 3} T^2\gg \alpha_s eB$ is a dynamic screening scale coming from the gluon self-energy that is discussed in Appendix \ref{appendix2}. }This has a logarithmic IR enhancement from the region
\be
\tilde\Lambda_{IR}\ll |p_z'-p_z|\ll {\changed{{\Lambda}_{\text{IR}}}\over m_q}E_p\ll E_p\,,
\ee
where we will get back to the other IR cutoff \changed{$\tilde\Lambda_{IR}$} shortly, and the last inequality is from our assumption $m_q^2\gg m_{D,B}^2\sim \alpha_s eB$. In this regime, we have legitimate approximations
\be
 |E_pp_z'-E_{p'}p_z|={m_q^2|p_z'^2-p_z^2|\over  |E_pp_z'+E_{p'}p_z|}\approx {m_q^2\over E_p}|p_z'-p_z|\,,
 \ee
and 
\be
2(E_pE_{p'}-p_zp_z'-m_q^2))={2m_q^2(p_z'-p_z)^2\over E_pE_{p'}+p_zp_z'+m_q^2)}\approx {m_q^2\over E_p^2}(p_z'-p_z)^2 \ll\changed{{\Lambda}^2_{\text{IR}}}\,,
\ee
so that we have in leading log order,
\bear
\gamma_q^{q-q}&\approx& 
8\pi\alpha_s^2T_RC_2(R)\left(eB\over 2\pi\right){m_q^2\over \changed{{\Lambda}^2_{\text{IR}}} E_p} f^{eq}_F(E_{p})(1- f^{eq}_F(E_{p}))
\int^{{\Lambda_{\text{IR}}\over m_q}E_p}_{\tilde\Lambda_{IR}}{dp_z'\over (2\pi)}{1\over |p_z'-p_z|}\nonumber\\
&=& 8\pi\alpha_s^2T_RC_2(R)\left(eB\over 2\pi\right){m_q^2\over \changed{{\Lambda}^2_{\text{IR}}} E_p} f^{eq}_F(E_{p})(1- f^{eq}_F(E_{p}))
\log\left({\Lambda}_{\text{IR}} E_p\over m_q \tilde\Lambda_{IR}\right)\,.\label{qdamp2211}
\eear
To identify the IR cutoff $\tilde\Lambda_{IR}$, we note that this IR divergence is from the Jacobian of the energy $\delta$-function which results \changed{in} the term $1/ |E_pp_z'-E_{p'}p_z|$ in (\ref{qdamp221}) (see (\ref{jac}) in Appendix \ref{appendix2}). This energy $\delta$-function will be smoothened precisely by the damping rate. From (\ref{qdamp2211}), we will see that $\gamma_q^{q-q}\ll \gamma_q^{1-2}\sim\alpha_s m_q^2/T$ in (\ref{qdamp12}). Therefore we can use in leading order,
\be
\tilde\Lambda_{IR}\sim \alpha_s m_q^2/T\,,
\ee
and we finally have (note that $\tilde\Lambda_{IR}\ll {{\Lambda}_{\text{IR}}\over m_q}E_p=\Lambda_{UV}$ is satisfied due to $eB\gg (T^2,m_q^2)$)
\bear
\gamma_q^{q-q}&=&\changed{8\over 3}\pi\alpha_s^2T_RC_2(R)\left(eB\over 2\pi\right){m_q^2\over \changed{{\Lambda}^2_{\text{IR}}} E_p} f^{eq}_F(E_{p})(1- f^{eq}_F(E_{p}))
\log\left(T^9eB\over \alpha_s^2 m_q^{11}\right)\nonumber\\&\sim& \changed{\alpha_s (\alpha_s eB m_q)^{1\over 3}\left({m_q\over T}\right)^3 \log\left(T^9eB\over \alpha_s^2 m_q^{11}\right) }\,.\eear
One can easily check that the UV regime $|p_z'-p_z|> {\Lambda_{IR}\over m_q}E_p$ produces a finite integral which adds
a constant under the log. \changed{This $\gamma_q^{q-q}$ is smaller than $\gamma_q^{1-2}\sim\alpha_s m_q^2/T$ in (\ref{qdamp12}).}

Lastly, let us compute the quark damping rate arising from 2-to-2 quark-gluon t-channel scattering, the collision term of which is worked out in Appendix \ref{appendix2}. We can start from the collision integral (\ref{semifin}) with the matrix element (\ref{apf1}) and the phase space integral (\ref{apf2}), which gives
\bear
\gamma_{q}^{q-g}&=&{8\pi^3\over 3}\alpha_s^2 N_c C_2(R) T^3{1\over E_p^4}\int {d^3{\bm q}\over (2\pi)^3}{1\over |{\bm q}|}
{1\over \left({\bm q}_\perp^2+{m_q^2\over E_p^2}q_z^2+\changed{{\Lambda}^2_{\text{IR}}}\right)^2}\nonumber\\&\times&\left(2m_q^4+5m_q^2p_z^2\left(1-{q_z^2\over |{\bm q}|^2}\right)+3p_z^4\left(1-{q_z^2\over |{\bm q}|^2}\right)^2\right)\,.
\eear
After changing a variable $q_z\to {E_p\over m_q}q_z$, and performing radial $|{\bm q}|$ integral, we finally obtain
\be
\gamma_{q}^{q-g}={\pi\over 3}\alpha_s^2 N_c C_2(R) {T^3\over \changed{{\Lambda}^2_{\text{IR}}}} \tilde S(m_q/E_p)\,,
\ee
where the dimensionless function $\tilde S(x)$ is defined by an angular integral
\bear
\tilde S(x)&=&x^4\int^1_{-1}d\cos\theta 
{1\over\sqrt{x^2\sin^2\theta+\cos^2\theta}}\\&\times&\left(2+5(1-x^2){\sin^2\theta\over (x^2\sin^2\theta+\cos^2\theta)}+3(1-x^2)^2{\sin^4\theta\over (x^2\sin^2\theta+\cos^2\theta)^2}\right)\,.\nonumber
\eear
In small $x=m_q/E_p\to 0$ limit, the angular integral localizes around $\cos\theta\sim x$ and parametrizing $\cos\theta=xt$, we have
\be
\tilde S(x)\to \int^\infty_{-\infty} dt {3\over (1+t^2)^{5/2}}=4\,,\quad x\to 0\,.
\ee
From \changed{${\Lambda}^2_{\text{IR}}\sim (\alpha_s eB/m_q^2)^{2\over 3} T^2$, we see that $\gamma_q^{q-g}\sim \alpha^2_s T(m_q^2/\alpha_seB)^{2/3} \ll \gamma_q^{1-2}\sim \alpha_s m_q^2/T$ when $eB\gg \alpha^{1/2}_s T^3/m_q$.
We will consider only such case in this section.}

\subsection*{Gluon damping rate}

We will consider the three important processes: 1) 1-to 2 process, 2) 2-to-2 gluon-quark t-channel scattering, and 3) 2-to-2 gluon-gluon t-channel scattering.

For the 1-to-2 process, we can start from the 1-to-2 collision term such as (\ref{collision}), but since we are considering the collision term for the gluons, we should replace the ${\bm k}$ integration
of gluon momentum with the ${\bm p}$ integration of incoming quark, and with a couple of changes of color and normalization factors, we have
\bear
\gamma_g^{1-2}&=&{1\over 2E_k (1+f^{eq}_B(E_k))} \int_{\bm p}\int_{{\bm p}'}|{\cal M}|^2(2\pi)^2\delta^{(2)}({\bm p}+{\bm p}'-{\bm k})(2\pi)\delta(E_p+E_{p'}-E_k)
\nonumber\\&\times& (1-f^{eq}_F(E_p))(1-f^{eq}_F(E_{p'}))\,,\eear
where the matrix element can be borrowed from (\ref{mat}),
\be
|{\cal M}|^2=2g_s^2 T_R e^{-{{\bm k}_\perp^2\over 2eB}} {{\bm k}_\perp^2\over |{\bm k}|^2}(E_pE_{p'}+p_zp_z'+m_q^2)\,.
\ee
Performing ${\bm p}'$ integration we have
\bear
\gamma_g^{1-2}&=&{1\over 2E_k (1+f^{eq}_B(E_k))}\left(eB\over 2\pi\right) \int {dp_z\over (2\pi)}{1\over (2E_p)(2E_{k-p})} |{\cal M}|^2(2\pi)\delta(E_p+E_{k-p}-E_k)
\nonumber\\&\times& (1-f^{eq}_F(E_p))(1-f^{eq}_F(E_{k-p}))\,.\eear
After some algebra the energy $\delta$-function becomes
\be
\delta(E_p+E_{k-p}-E_k)={2 E_p E_{k-p}\over {\bm k}_\perp^2\sqrt{1-4m_q^2/{\bm k}_\perp^2}}\delta(p_z-p_z^\pm)\Theta({\bm k}_\perp^2-4m_q^2)
\,,
\ee
where 
\be
p_z^\pm= {k_z\over 2}\pm {E_k\over 2}\sqrt{1-4m_q^2/{\bm k}_\perp^2}\,,
\ee
and the matrix element simply becomes 
\be
|{\cal M}|^2=4g_s^2 T_R {m_q^2}\,.
\ee
Performing $p_z$ integral, we finally have (for ${\bm k}_\perp^2 >  4m_q^2$)
\bear
\gamma_{g}^{1-2}&=&{8\pi\alpha_s T_R\over E_k (1+f^{eq}_B(E_k))}\left(eB\over 2\pi\right)
\left(m_q^2\over {\bm k}_\perp^2\sqrt{1-4m_q^2/{\bm k}_\perp^2}\right)(1-f^{eq}_F(E_p))(1-f^{eq}_F(E_{k-p}))\nonumber\\
&\sim& \alpha_s m_q^2/T \left(eB\over T^2\right)\,,\label{dampg12}
\eear
where
\be
(E_p, E_{k-p})={1\over 2}\left(E_k\pm k_z\sqrt{1-4m_q^2/{\bm k}_\perp^2}\right)\,.
\ee
The above damping rate exists only for ${\bm k}_\perp^2> 4m_q^2$, which is okay for our purpose since we are interested in
the gluonic contribution to the transverse color conductivity that comes mainly from the hard gluons moving transverse to the magnetic field, so that $m_q^2\lesssim {\bm k}_\perp^2\sim T^2$.

Next, let's consider 2-to-2 gluon-quark/antiquark scattering contributions. Again, we can start from the collision term similar to the case of quark-gluon scattering in (\ref{2-2glu}) with the matrix element (\ref{apf1}),
\bear
\gamma_g^{g-q}&=&{2\over 2E_k(1+f^{eq}_B(E_k))}\int_{\bm k'}\int_{\bm p}\int_{\bm p'}|{\cal M}|^2 (2\pi)^2\delta({\bm k}+{\bm p}-{\bm k'}-{\bm p'})(2\pi)\delta(E_k+E_p-E_{k'}-E_{p'})\nonumber\\
&\times& (1+f_B^{eq}(E_{k'}))f_F^{eq}(E_p)(1-f_F^{eq}(E_{p'}))\,,
\eear
where the factor 2 comes from equal contributions from quarks and antiquarks, and 
\be
|{\cal M}|^2=32 g_s^4N_c T_R {(E_kE_p-k_zp_z)^2\over ({\bm q}_\perp^2+{m_q^2\over E_p^2}q_z^2+\changed{{\Lambda}^2_{\text{IR}}})^2}\,.
\ee
Changing integration variable from $\bm k'$ to ${\bm q}={\bm k}-{\bm k'}$ and performing ${\bm p'}$ integration, we have in small $q/T$-limit as
\bear
\gamma_g^{g-q}&=&{1\over E_k(1+f^{eq}_B(E_k))}\left(eB\over 2\pi\right)\int{d^3{\bm q}\over (2\pi)^3 2E_{k-q}}\int{dp_z\over (2\pi)} {|{\cal M}|^2\over (2E_p)(2E_{p-q})}\nonumber\\
&\times&(2\pi)\delta \left(\hat{\bm k}\cdot {\bm q}-{p_z\over E_p}q_z\right)(1+f_B^{eq}(E_{k-q}))f_F^{eq}(E_p)(1-f_F^{eq}(E_{p-q}))\,.
\eear

Since we are interested in the gluons moving transverse to the magnetic field, let us focus on the case $k_z=0$ which simplifies the computation, and take ${\bm k}=k \hat x$. Performing $q_x$ integration with the energy $\delta$-function, we finally have
\bear
\gamma_g^{g-q}&=& 4g_s^4 N_c T_R \left(eB\over 2\pi\right)\int {dq_y dq_z\over (2\pi)^2} {1\over (q_y^2+q_z^2+\changed{{\Lambda}^2_{\text{IR}}})^2}\int {dp_z\over (2\pi)} f_F^{eq}(E_p)(1-f^{eq}_F(E_p))\nonumber\\
&=&16\alpha_s^2 N_c T_R\left(eB\over 2\pi\right){T\over \changed{{\Lambda}^2_{\text{IR}}}}\int_0^\infty { dp_z\over T} f_F^{eq}(E_p)(1-f^{eq}_F(E_p))\nonumber\\&
\sim& \changed{\alpha_s (\alpha_s eB m_q)^{1\over 3} \left({m_q\over T}\right)} \,.
\label{eq:g_g-q}
\eear
We see that $\gamma_q^{g-q}$ is smaller than $\gamma_g^{1-2}$ in (\ref{dampg12}).

Finally, we discuss the contribution from the 2-to-2 gluon-gluon t-channel processes. 
For this we don't need to compute since it is the same process that gives the dominant damping rate in the usual plasma, 
and need only to discuss the screening mass. 
As explained below Eq.~(\ref{eq:Gprop}), there are two polarizations of the exchanged transverse gluons 
which are, respectively, screened by the self energy from the 1-loop LLL quarks $m_{D,B}^2\sim\alpha_s eB$, or by that from the 1-loop thermal gluons $m_{D,T}^2\sim\alpha_s T^2$. 
Explicitly, the latter mode has the polarization that is perpendicular to the plane formed by the gluon momentum and $\bm B$ field, and is decoupled from the 1+1 dimensional LLL quarks.
Therefore, when $m_{D,B} \gg m_{D,T}$, the dominant contribution comes from 
the exchange of the gluon with this latter polarization which is screened only by $ m_{D,T}$. 
Inserting $ m_{D,T}$, we get 
 \be
 \label{eq:damping-gg}
 \gamma_g^{g-g}\sim \alpha_s^2 {T^3\over m_{D,T}^2}\log\left(m_{D,T}\over\alpha_s T\right)
 \sim \alpha_s T \log \alpha_s^{-1}\,,
\ee
where $T^3$ in the numerator is from the thermal gluon density,
while $m_{D,T}^2$ in the denominator is from the screening scale in the t-channel propagator.


In summary of the above computations, the dominant damping rates for quarks and gluons are

1) When $m_q^2/T^2\gg T^2/eB$,
\be
\gamma_q\sim  \alpha_s  (m_q^2/T)\,,\quad \gamma_g\sim\alpha_s  m_q^2/T \left(eB\over T^2\right)\,.
\ee

2) When $m_q^2/T^2\ll T^2/eB$,
\be
\gamma_q\sim \changed{ \alpha_s  (m_q^2/T)} \,,\quad \gamma_g\sim  \alpha_s T \,.
\ee
As claimed before, the quark damping rate is parametrically smaller than the gluon damping rate by $T^2/eB\ll 1$ or \changed{by $m_q^2/T^2\ll 1$}.
The resulting color conductivity from (\ref{colorc}) is

1) When $m_q^2/T^2\gg T^2/eB$,
\be
\sigma_c^{L}\sim {(eB) T\over  m_q^2}\,,\quad\sigma_c^T\sim
{ T^5\over  m_q^2 eB}\,.
\ee

2) When $m_q^2/T^2\ll T^2/eB$,
\be
\sigma_c^{L}\sim \changed{{(eB) T\over  m_q^2}}\,,\quad\sigma_c^T\sim
{T}\,.
\ee

In the physics of sphaleron transitions, the typical length scale is given by the magnetic scale $l_{sph}^{-1}=k\sim \alpha_s T$ while the time scale is governed by the magnetic diffusion time (Lenz's law) (\ref{magdif}), $t_{sph}^{-1}\sim k^2/\sigma_c\sim \alpha_s^2 T^2/\sigma_c$, so the sphaleron transition rate scales as $\Gamma_s\sim (l_{sph}^{-1})^3 t_{sph}^{-1}\sim \alpha_s^5 T^5/\sigma_c$ \cite{Arnold:1998cy}.
The increased $\sigma_c$ along the magnetic field (while the transverse color conductivity remains similar) would therefore reduce the transition rate. However, the $\sigma_c$ to be used in this estimate should be the one defined at the spatial scale $k\sim \alpha_s T$, and if the mean free path (equivalently, the inverse damping rate $\gamma^{-1}$) that gives the above results for color conductivity is larger than this spatial scale, we need to use $k^{-1}\sim (\alpha_s T)^{-1}$ instead as the effective mean free path to determine
the $\sigma_c$ used in the estimate for sphaleron transitions \cite{Arnold:1998cy}.
Considering this fact, we find that the effective longitudinal $\sigma_c^L$ to be used for sphaleron transitions becomes in $m_q\ll T$ case
\be
\sigma_c^L(k\sim\alpha_s T)\sim {eB\over T} \gg T\,,
\ee
while the transverse color conductivity is
\bear
\sigma_c^T(k\sim\alpha_s T)&\sim& {T^5\over m_q^2 eB}\lesssim T\quad  ({\rm when }\,\,m_q^2/T^2\gtrsim T^2/eB)\,,\nonumber\\
\sigma_c^T(k\sim\alpha_s T)&\sim&  T\quad  \quad\quad\quad \quad({\rm when }\,\,m_q^2/T^2\ll T^2/eB)\,.
\eear
This $\sigma_c^L$ is much larger than the usual value $T/\log(1/\alpha_s)$, while $\sigma_c^T$ remains similar to that. It means that the sphaleron transition rate will be smaller in the presence of strong magnetic field due to the enhanced Lenz's law in color field dynamics, as claimed in the introduction.

\subsection*{Acknowledgment}

We thank Mark Alford for helpful comments and inputs on quark mass dependence, and also thank Pavel Buividovich, Chuck Horowitz, Wai-Yee Keung, Sanjay Reddy, Dirk Rischke, Thomas Schaefer, Armen Sedrakian, and Misha Stephanov for discussions.
This work is supported in part by China Postdoctoral Science Foundation under Grant No.~2016M590312 
and Japan Society for the Promotion of Science Grants-in-Aid No.~25287066 (K.H.), by
the U.S. Department of Energy, Office of Science, Office of Nuclear Physics, within the framework of the Beam Energy Scan Theory (BEST) Topical Collaboration (S.L. and H.U.Y.), by the Alexander von Humboldt Foundation (D.S.).
H.U.Y. thanks the INT program "The Phases of Dense Matter" where part of this work was performed.
We appreciate the hospitality and support provided by RIKEN-BNL Research Center and Institute for Nuclear Theory, University of Washington.

\appendix

\section{Feynman rules in the LLL approximation} 

In this appendix, we summarize the quantization of quark field in the presence of a strong background magnetic field
and derive the effective Feynman rules that we use in computing the collision terms in the Boltzmann equation.
We choose to work in the Landau gauge $A^2=B x^1$ (with $\bm B=B \hat x^3\equiv B\hat z$) which seemingly breaks the translational invariance in $x^1$ direction, while keeping that in $x^2$ direction. This allows us to introduce two momentum
quantum numbers, $p_z$ and $p_2$, along $\hat z$ and $\hat x^2$. It is important to keep in mind that 1) there is no concept of $p_1$ in the quark wave functions (while the gluon wave functions have it), and 2) $p_2$ serves as a label for the degenerate Landau levels in the transverse $(x^1,x^2)\equiv {\bm x}_\perp$ space.

To take care of the transverse density of states of the Landau levels in a clear manner, we first consider a finite box of
each sides $(L_1, L_2, L_3)$, and then take an infinite volume limit at the end.
The two dimensional momenta $(p_z,p_2)$ take discrete values, which we denote collectively as $\bm p_n$.
Solving the Dirac equation with the background magnetic field, we get positive/negative energy solutions as usual,
\be
e^{-i E_{n,l} x^0+i{\bm p}_n\cdot\bm x} u^{l}_\parallel(p_z) {\cal H}_l\left(x^1-{p_2\over eB}\right)\,,\quad
e^{+i E_{n,l} x^0-i{\bm p}_n\cdot\bm x} v^{l}_\parallel(p_z) {\cal H}_l\left(x^1+{p_2\over eB}\right)\,,
\ee
where the energy is
\be
E_{n,l}=\sqrt{p_z^2+(2l+1\mp 1)|eB|+m_q^2}\equiv \sqrt{p_z^2+m_{l}^2}\,,
\ee
depending on the spinor projection $i\gamma^1\gamma^2=\pm 1$, and ${\cal H}_l(x^1)$
are the normalized $l$-th eigenstate of simple harmonic oscillator with frequency $\omega=|eB|$, such that
\be
{\cal H}_0(x^1)=\left(|eB|\over \pi\right)^{1\over 4} \exp\left(-{(x^1)^2\over 2|eB|}\right)\,,
\ee
and $u^l_\parallel(p_z)$ and $v^l_\parallel(p_z)$ are 1+1 dimensional spinors (due to the projection $i\gamma^1\gamma^2=\pm 1$) for quarks and antiquarks with the mass $m_l$ in relativistic normalization $u^\dagger u=v^\dagger v=2E_{n,l}$. It is important to notice that the quark state with $p_2$ is localized
in $x^1\sim {p_2/eB}$ with a width $\Delta x^1\sim 1/\sqrt{|eB|}$, while the antiquark state is localized in $x^1\sim -{p_2/eB}$.

Let us first reproduce the well-known transverse density of states of Landau levels in this gauge: $|eB|/(2\pi)$.
The $p_2$ is discrete valued, $p_2=2\pi k/L_2$, with integers $k$, and the quark state with a given Landau level $l$ with this momentum is localized 
in $x^1\sim {p_2/eB}=2\pi k/(L_2 eB)$. Since $x^1$ should lie in the interval $[0,L_1]$, we have $0< k< L_1L_2(|eB|/2\pi)$,
that is, the total number of such states is $L_1L_2(|eB|/2\pi)$ per the transverse area $L_1 L_2$.

The special case with $l=0$ and $i\gamma^1\gamma^2=+1$ gives the lowest possible 1+1 dimensional mass $m_0^2=m_q^2$, which is separated by multiples of $|eB|$ from other higher level states. These states are the lowest Landau levels (LLL). In the language of 1+1 dimension, their spinors $u^0_\parallel(p_z)$, $v^0_\parallel(p_z)$ form
a {\it single} Dirac fermion field in 1+1 dimension.
For higher Landau levels with $m_l^2=2l|eB|+m_q^2$ ($l\ge 1$), we have two possibilities to get the same mass: one with $i\gamma^1\gamma^2=+1$ and the level $l$, and the other with $i\gamma^1\gamma^2=-1$ and the level $l-1$.
These two possibilities result in {\it two} Dirac fermion fields with the common mass $m_l$ in the language of 1+1 dimensions.

Following the standard quantization scheme, we expand the quark field operator as
\be
\psi(\bm x)={1\over \sqrt{L_2L_3}}\sum_{{\bm p}_n,l}{1\over \sqrt{2 E_{n,l}}} \left(e^{i{\bm p}_n\cdot{\bm x}}{\cal H}_l\left(x^1-{p_2\over eB}\right) u^l_\parallel(p_z) a_{{\bm p}_n,l}+e^{-i{\bm p}_n\cdot{\bm x}}{\cal H}_l\left(x^1+{p_2\over eB}\right) v^l_\parallel(p_z) b^\dagger_{{\bm p}_n,l}\right)\,,
\ee
where the sum over $i\gamma^1\gamma^2=\pm 1$ is assumed, and 
\be
\{ a_{{\bm p}_n,l},a^\dagger_{{\bm p}_{n'},l'}\}=\{ b_{{\bm p}_n,l},b^\dagger_{{\bm p}_{n'},l'}\}=\delta_{n,n'}\delta_{l,l'}\,.
\ee
Using the completeness relation 
\be
\sum_l {\cal H}_l(x){\cal H}_l(y)=\delta(x-y)\,,
\ee
it is easy to show the canonical commutation relation is satisfied
\be
\{\psi_\alpha(\bm x),\psi^\dagger_\beta(\bm y)\}=\delta^{(3)}({\bm x}-{\bm y})\delta_{\alpha\beta}\,.
\ee

Since the higher Landau level states have the energy at least of order $\sqrt{|eB|}$, their thermal occupation numbers
are exponentially small $e^{-\sqrt{|eB|}/T}$ in our assumed hierarchy $T^2\ll eB$, and they don't contribute to the transport coefficients such as electric conductivity of our interest. This justifies the LLL approximation that we use in this work, that is
keeping only $l=0$ and $i\gamma^1\gamma^2=+1$ component in the above expansion of quark field operator.
In the following, we will call the LLL spinors $(u^0_\parallel(p_z),v^0_\parallel(p_z))$ simply by $(u(p_z),v(p_z))$, and
similarly $E_{n,0}\equiv E_n$ and $a_{{\bm p}_n,0}\equiv a_{{\bm p}_n}$, so that
\be
\psi(\bm x)\sim{1\over \sqrt{L_2L_3}}\sum_{{\bm p}_n}{1\over \sqrt{2 E_{n}}} \left(e^{i{\bm p}_n\cdot{\bm x}}{\cal H}_0\left(x^1-{p_2\over eB}\right) u(p_z) a_{{\bm p}_n}+e^{-i{\bm p}_n\cdot{\bm x}}{\cal H}_0\left(x^1+{p_2\over eB}\right) v(p_z) b^\dagger_{{\bm p}_n}\right)\,.
\ee
One consequence of the LLL approximation is that the quark current $j^\mu=\bar\psi\gamma^\mu\psi$ has
zero component in the transverse $\bm x_\perp$ direction, due to the projection $i\gamma^1\gamma^2=+1$ which anti-commutes with $\gamma^\perp$. Physically this is because Landau level states move only along 1+1 dimensions.
The transverse current or transverse motion necessarily involves mixing with higher Landau levels.

We are interested in the QCD interaction with the gluon fields living in 3+1 dimensions. 
We will do time-ordered perturbation theory, but the matrix element for a given Feynman diagram 
ends up to a (1+1 dimensional) relativistic expression after summing over all time-ordered processes. We will derive such Feynman rules
in our LLL approximation by showing a few example time-ordered perturbation theory computations and extracting Feynman rules from those results.

The interaction Hamiltonian is
\be
H_I=g_s \int d^3\bm x\,\,A^a_\mu(\bm x)\bar\psi(\bm x)\gamma^\mu_\parallel t^a \psi(\bm x)\,,\label{Hi}
\ee
where $a$ is the color index, and recall that in the LLL approximation $\mu$ runs only along 1+1 dimensions indicated by $\gamma^\mu_\parallel$. Since $\psi$ field is already projected by $i\gamma^1\gamma^2=+1$, the $\gamma^\mu_\parallel$ matrices are effectively $2\times 2$ $\gamma$ matrices in 1+1 dimensions.
The gluon field is quantized as usual:
\be
A_\mu(\bm x)={1\over \sqrt{V}}\sum_{{\bm q}_m,\epsilon} {1\over \sqrt{2 |{\bm q}_m|}} e^{i{\bm q}_m\cdot\bm x}\epsilon_\mu\, 
a^g_{{\bm q}_m}+{\rm h.c.}\,,
\ee
where $V=L_1L_2L_3$ and ${\bm q}_m$ is the discrete 3-momentum, and $[a_{{\bm q}_m},a^\dagger_{{\bm q}_{m'}}]=\delta_{m,m'}$.
The $H_I$ has non-zero matrix elements for four types of processes: absorption/emission of a gluon by/from quark or antiquark, and pair creation/annihilation of quark-antiquark pair from/to a gluon.
For example, denoting a normalized one quark state as $|{\bm p}_{n'}\rangle$, and one quark+one gluon state as $|{\bm p}_{n},{\bm k}_m\rangle$, we have
\be
\langle {\bm p}_{n'}| H_I | {\bm p}_{n},{\bm k}_m\rangle={g_s\over\sqrt{V}}{1\over\sqrt{2E_n}}{1\over\sqrt{2E_{n'}}}{1\over\sqrt{2|{\bm k}_m|}} \epsilon_\mu \left(\bar u(p_z')\gamma^\mu_\parallel u(p_z)\right) R_{00}({\bm k}_{m\perp}) e^{i\Sigma}\delta^{(2)}_{{\bm p}_n+{\bm k}_m-{\bm p}_{n'}}
\ee
where $\delta^{(2)}$ is only about ${\bm p}_n=(p_z,p_2)$ (so that $k^1_{m}$ is not constrained at all), and the form factor $R_{00}({\bm k}_{m\perp})$ and the Schwinger phase $e^{i\Sigma}$ arise from the overlap integral
\be
\int dx^1\, e^{i{\bm k}^1_m x^1} {\cal H}_0\left(x^1-{p_2/eB}\right) {\cal H}_0\left(x^1-{p'_2/eB}\right)
= R_{00}({\bm k}_{m\perp}) e^{i\Sigma}\,,
\ee
with
\be
R_{00}({\bm k}_{\perp})=e^{-{{\bm k}_{\perp}^2\over 4|eB|}}\,,\quad \Sigma=-{k^1_m\over 2eB}\left(p_2+p_2'\right)\,.
\ee
Note that we used the fact that $k_m^2=p_2'-p_2$ in the expression of $R_{00}({\bm k}_{m\perp})$.
The other matrix elements of $H_I$ are similar with the form factor and the Schwinger phase. From these and applying the Fermi's Golden rule, we can construct
the collision term in the Boltzmann equation as a transition probability rate per unit time from a given initial state to a final state. In this way, the normalization issue is taken care of clearly in a finite volume we are considering before we take an infinite volume limit.

As a first example, let us consider the collision term for the quark distribution of momentum ${\bm p}_n$ from 2-to-2 quark scattering: ${\bm p}_n+{\bm p}_{n''}\to {\bm p}_{n'}+{\bm p}_{n'''}$. There are two time ordered diagrams in the second order perturbation theory where the transition rate is given by
\be
T_{i\to f}=\sum_{m}{\langle f|H_I|m\rangle\langle m |H_I|i\rangle\over {E_m-E_i}}(2\pi)\delta(E_f-E_i)\,,
\ee
Summing the two time ordered processes, for which $|m\rangle=|{\bm p}_{n'},{\bm p}_{n''},{\bm q}_m\rangle$ or
$|m\rangle=|{\bm p}_{n},{\bm p}_{n'''},{\bm q}_m\rangle$ (${\bm q}_m$ is the exchanged gluon momentum), we get after a short algebra
\bear
\sum_{n',n'',n'''}T_{i\to f}&=& {1\over (L_2L_3)^2}\sum_{n',n'',n'''}{1\over 2E_n}{1\over 2E_{n'}}{1\over 2E_{n''}}{1\over 2E_{n'''}}\delta^{(2)}_{{\bm p}_n+{\bm p}_{n''}-{\bm p}_{n'}-{\bm p}_{n'''}}\nonumber\\
&\times& \left|{\cal M}\right|^2 (2\pi)\delta\left(E_n+E_{n''}-E_{n'}-E_{n'''}\right)\,,
\eear
where the matrix element is given by
\be
{\cal M}=g_s^2{1\over L_1}\sum_{{\bm q}^1_m}{\eta_{\mu\nu}\over (q_m^0)^2-{\bm q}_m^2} \left(R_{00}({\bm q}_{m\perp})\right)^2 e^{-i{q^1_m\over 2eB}(p_2+p_2'-p_2''-p_2''')}[\bar u(p_z')\gamma^\mu u(p_z)][\bar u(p_z''')\gamma^\nu u(p_z'')]\,,
\ee
with $q^0_m=E_{n'}-E_n$ and ${\bm q}_m^{(2)}={\bm p}_n-{\bm p}_{n'}$. The structure of ${\cal M}$ is a product of 1+1 dimensional relativistic matrix element for quarks and the form factors/Schwinger phase. The gluon propagator is 3+1 dimensional. Recall that there is no constraint for ${\bm q}_m^1$ and we have a summation over it in ${\cal M}$.
We omit color factors in the above and the following, but can easily be reinstated.
The above transition probability rate with distribution functions of incoming and outgoing states attached is what should
appear in the collision term in the Boltzmann equation.
Taking an infinite volume limit, we get a collision term
\bear
C[f_+({\bm p})]&=&-{1\over 2E_{p}}\int_{\bm p'}\int_{\bm p''}\int_{\bm p'''} |{\cal M}|^2 (2\pi)^2\delta^{(2)}\left({\bm p}+{\bm p''}-{\bm p'}-{\bm p'''}\right)\\ &\times&(2\pi)\delta\left(E_{p}+E_{p''}-E_{p'}-E_{p'''}\right)
f_+({\bm p})f_+({\bm p''})(1-f_+({\bm p'}))(1-f_+({\bm p'''}))\,,\nonumber
\eear
with
\be
{\cal M}=g_s^2\int{d{\bm q}^1\over (2\pi)}{\eta_{\mu\nu}\over q^2 } \left(R_{00}({\bm q}_{\perp})\right)^2 e^{-i{q^1\over 2eB}(p_2+p_2'-p_2''-p_2''')}[\bar u(p_z')\gamma^\mu u(p_z)][\bar u(p_z''')\gamma^\nu u(p_z'')]\,,\label{m}
\ee
and 
\be
\int_{\bm p}\equiv \int{dp_z dp_2 \over  (2\pi)^2 2E_{p}}\,.
\ee
One can work out to see that the Schwinger phase in (\ref{m}) is crucial to get a finite result with correct Landau level density of state. 

As another example, let's consider 2-to-2 scattering of a quark with thermal gluons: ${\bm p}+{\bm k''}\to {\bm p'}+{\bm k'''}$.
Working out similar details as above and taking an infinite volume limit, one arrives at
\bear
C[f_+({\bm p})]&=&-{1\over 2E_{p}}\int_{\bm p'}\int_{\bm k''}\int_{\bm k'''} |{\cal M}|^2 (2\pi)^2\delta^{(2)}\left({\bm p}+{\bm k''}-{\bm p'}-{\bm k'''}\right)\\ &\times&(2\pi)\delta\left(E_{p}+E_{k''}-E_{p'}-E_{k'''}\right)
f_+({\bm p})f_g({\bm k''})(1-f_+({\bm p'}))(1+f_g({\bm k'''}))\,,\nonumber
\eear
where
\be
{\cal M}=g_s^2 f^{abc}R_{00}({\bm q}_\perp)e^{-i{{\bm q}^1\over 2eB}(p_2+p_2')} {\eta_{\mu\nu}\over q^2}[\bar u(p_z')\gamma^\mu u(p_z)]\times ({\rm gluon\,\,current})\,,
\ee
is the usual relativistic expression except the form factor/Schwinger phase, and
\be
\int_{\bm k}\equiv \int{d^3{\bm k}\over (2\pi)^3 2E_k}\,.
\ee

As a final example, let's consider 2-to-2 scattering of a gluon with thermal LLL quarks: ${\bm k}+{\bm p''}\to {\bm k'}+{\bm p'''}$ (the same process as the second example, but the collision term for the gluon tagged, rather than the quark).
Taking an infinite volume limit, we end up to
\bear
C[f_g({\bm k})]&=&-{1\over 2E_{k}}{1\over L_1}\int_{\bm k'}\int_{\bm p''}\int_{\bm p'''} |{\cal M}|^2 (2\pi)^2\delta^{(2)}\left({\bm k}+{\bm p''}-{\bm k'}-{\bm p'''}\right)\\ &\times&(2\pi)\delta\left(E_{k}+E_{p''}-E_{k'}-E_{p'''}\right)
f_g({\bm k})f_+({\bm p''})(1+f_g({\bm k'}))(1-f_+({\bm p'''}))\,,\nonumber
\eear
Note the residual $1/L_1$ factor which is correct as we explain in the following.
In this case, when we sum over the final quark states with $p''$ and $p'''$, one easily see that $p_2''+p_2'''$ is unconstrained, and one has a trivial summation over them. The physics is simple to understand: recalling that $x^1={p_2\over eB}$, the $(p'''+p''')/2\sim eB x^1_c$ represents a center of mass $x^1$ position of the incoming and out-going quark states, which is free to take any value between $(0,L_1)$. Indeed,
in taking an infinite volume limit, one encounters the combination
\be
{1\over L_1}{1\over 2\pi}\int d(p_2''+p_2''')/2 \to {1\over L_1}(eB/2\pi)\int_0^{L_1} dx^1_c= (eB/ 2\pi)\,,
\ee
that is, the unconstrained integral of $(p_2''+p_2''')/2$ always comes with a residual $L_1$ factor in the denominator,
and results in the transverse density of states of LLL, $(eB/2\pi)$. This is generic for any complete fermion line whose
phase space is integrated: there is one $p_2$ integral associated to it that is not constrained at all (which represents the overall $x^1$ position of the fermions), and it always comes with a residual $1/L_1$ factor to produce $(eB/2\pi)$ at the end.
Note that this rule does not apply for the tagged fermion line in the collision term as in the second example, since the tagged
fermion momentum is not integrated over.

From these examples, one derives the following Feynman rules in the LLL approximation:
1) For external quark/antiquark lines, the phase space integration is 
\be
\int_{\bm p}\equiv \int{dp_z dp_2 \over  (2\pi)^2 2E_{p}}\,.
\ee
while for external gluons, it is
\be
\int_{\bm k}\equiv \int{d^3{\bm k}\over (2\pi)^3 2E_k}\,.
\ee

2) For quark-quark-gluon vertex, impose the momentum conservation only along two dimensions $(p_z,p_2)$,
and attach the form factor and Schwinger phase. The $k_1$ component of gluon is not constrained.

3) If there is an internal ${\bm q}^1$ gluon momentum which is not fixed by external gluons, we integrate $\int d{\bm q}^1/(2\pi)$ in the total matrix element ${\cal M}$.

4) The rest of the matrix element simply follows the usual relativistic Feynman rules for 1+1 dimensional relativistic fermions and 3+1 dimensional relativistic gauge theory.

5) In the collision integral, the momentum $\delta$-function is only two dimensional, and the energy $\delta$-function is as usual. There is an overall normalization of $1/(2E_p)$ in front of the collision term.

6) There exists one unconstrained $p_2$ integral for any complete quark (antiquark) line whose phase space is integrated.
We have a simple thumb rule that each of these unconstrained $p_2$ integral produces the transverse density of states of LLL, $(eB/2\pi)$;
\be
\int {dp_2\over (2\pi)}\to \left(eB\over 2\pi\right).
\ee
In fact, only with this thumb rule applied, the final result has the correct energy dimension for the collision term.

\section{Collision terms from the 2-to-2 processes \label{appendix2}}

In this appendix, we give explicit derivations of some of the collision terms arising from 2-to-2 processes, and show that these
are indeed sub-leading compared to the 1-to-2 process in the main text when $\alpha_s eB\ll m_q^2$.
We will also see in passing that in the other regime of $m_q^2\ll \alpha_s eB$, some of these 2-to-2 processes become of the same order as the 1-to-2 process, as we claimed in the introduction.

There are several 2-to-2 processes, and we will illustrate that their largest contribution to the collision term when $\alpha_s eB\ll m_q^2$ is 
\be
C_{2-2}[\chi_+]\sim \alpha_s^2 \left({eB}\right)\log\left(\changed{m_q^5\over \alpha_s eB T^3}\right) \partial_{p_z}^2\chi_+\,,\label{2to2}
\ee
compared to the 1-to-2 collision term in the main text which is of order
\be
C_{1-2}[\chi_+]\sim \alpha_s \left(m_q^2\over T^2\right) \chi_+\,,\label{1to2app}
\ee
so that indeed 2-to-2 is subleading to 1-to-2 when $\alpha_s eB\ll m_q^2$.

The most important 2-to-2 processes are quark-antiquark scatterings.
One reason why they are dominant over quark-gluon scatterings is simply the thermal density of scattering particles: antiquark thermal density is $\sim (eB/2\pi) T$ which is bigger than the thermal density of gluons $\sim T^3$.
However, it would be still comforting to check this expectation explicitly, since quark and antiquark currents are only 1+1 dimensional,
and the corresponding matrix elements may depend on the quark mass in a non-trivial way. Indeed, we will see 
in the example of quark-antiquark t-channel scatterings that there
is an intricate cancellation of quark mass dependence in quark-antiquark scatterings that results in the estimate (\ref{2to2}).
We will also show an example computation of quark-gluon t-channel scattering contribution to confirm the expectation that
quark-gluon scatterings are indeed sub-leading compared to (\ref{2to2}). As in the case without magnetic field, the t-channel processes are potentially enhanced by an extra IR logarithm compared to the s-channel processes, so we will present explicit computations for the t-channel processes only.

The existence of on-shell 1-to-2 processes implies that the s-channel 2-to-2 scatterings have a on-shell singularity when the s-channel gluon becomes close to the on-shell point.
One might worry whether this enhanced contribution might overturn our power counting estimates claimed in the above. However, we explain in the following that this singular s-channel 2-to-2 contribution
is precisely what our 1-to-2 collision term (\ref{finalcol}) is.
The s-channel singularity is caused by a long-lived intermediate gluon, and in the narrow width approximation which is valid in leading order, it is regulated by a finite damping rate of gluon in the retarded gluon propagator that gives a finite life-time: the dominant damping rate is given by 1-to-2 process as
in (\ref{dampg12}). The resulting contribution to the collision term arising from the damping-rate regulated 2-to-2 s-channel singularity can be shown to be ``identical" to our 1-to-2 collision term (\ref{finalcol})\footnote{More precisely, one first solves the gluon Boltzmann equation with 1-to-2 collision term to express the gluon distribution in terms of \changed{quark and anti-quark distribution functions,} 
 and replace the gluon distribution in the 1-to-2 collision term for quark Boltzmann equation with that solution. The result is identical to the damping rate regulated s-channel 2-to-2 collision term.}, 
which means that this 2-to-2 contribution near s-channel singularity is in fact precisely taken care of by having our 1-to-2 collision term, that is, it would be a double counting to have them both.
The physics is clear: 1-to-2 collision assumes that the external gluon state has a narrow width (damping rate), so we can treat it as a stable particle. In reality, the gluon is not stable and will decay to quark-antiquark
final states eventually, so the full process should be 2-to-2 s-channel process. When the damping rate is narrow, the decay process will happen sufficiently long after the initial 1-to-2 process happens,
so the initial 1-to-2 process is factorized from the final decay processes. As the initial 1-to-2 process does't care what happens to the gluon afterwards, if we sum over all possible final states of s-channel 2-to-2 processes, the resulting rate should be the same to the initial 1-to-2 process rate. The same physics can be found in the Z-boson physics in the "narrow width approximation".
Outside the singular region, the off-shell s-channel contribution is parametrically smaller than (\ref{2to2}) by an absence of logarithm.

The readers might wonder why we don't care about quark-quark scatterings at all.
The reason is simply due to 1+1 dimensional kinematics of massive quasi-particles.
Imagine we consider quark-quark scatterings: $\bm p+\bm p''\to \bm p'+\bm p'''$.
Remembering that the dispersion relation is 1+1 dimensional, $E_p=\sqrt{p_z^2+m_q^2}$, the energy and $z$-momentum
conservation allows only two possibilities of final momenta $(p_z',p_z''')$: either $(p_z',p_z''')=(p_z,p_z'')$
or $(p_z',p_z''')=(p_z'',p_z)$, that is, the final $z$-momenta should be the same to the initial momenta up to permutation. The collision term, using detailed balance at equilibrium,
is proportional to 
\be
\chi_+(p_z')+\chi_+(p_z''')-\chi_+(p_z)-\chi_+(p_z'')=0\,,\label{555}
\ee
where the crucial point is that the distributions depend only on $p_z$, not $p_2$ due to homogeneity in the transverse 
plane in our specific problem for longitudinal conductivity. Therefore, the collision term from quark-quark scatterings 
for our specific problem vanishes simply due to kinematics. 
Physics wise, quark-quark scatterings leave two incoming $p_z$ momenta unchanged (only permuted), while the transverse positions parametrized by $p_2$ can be changed. For longitudinal conductivity, these transverse diffusion does not matter at all, and only $p_z$ that determines the longitudinal velocity matters: therefore, quark-quark scatterings are irrelevant for longitudinal conductivity.
Note that this wouldn't be the case if the source
is not homogeneous in the transverse plane (so that $\chi_+$ depends also on $p_2$).

The situation is different for quark-antiquark scatterings, where the collision term is proportional to 
\be
\chi_+(p_z')+\chi_-(p_z''')-\chi_+(p_z)-\chi_-(p_z'')\,.
\ee
The case of $(p_z',p_z''')=(p_z,p_z'')$ still gives vanishing result, but the other case $(p_z',p_z''')=(p_z'',p_z)$ (momenta permuted between the quark and the antiquark) gives
\be
\chi_+(p_z')+\chi_-(p_z''')-\chi_+(p_z)-\chi_-(p_z'')=2(\chi_+(p_z')-\chi_+(p_z))\,,\label{584}
\ee
where we used the fact $\chi_+(p_z)=-\chi_-(p_z)$. Note that this case has a non-zero momentum exchange $q=p'-p\neq 0$.

\subsection*{Quark-antiquark t-channel scattering}

The collision term from quark-antiquark scattering, $\bm p+\bm p''\to \bm p'+\bm p'''$, and its time reversed process is written as
\bear
C[f_+(p_z)]&=&{1\over 2E_p}\int_{\bm p'}\int_{\bm p''}\int_{\bm p'''}|{\cal M}|^2 (2\pi)^2\delta^{(2)}(\bm p+\bm p''-\bm p'-\bm p''')(2\pi)\delta(E_p+E_{p''}-E_{p'}-E_{p'''})\nonumber\\
&\times&\beta f^{eq}_F(E_p)f^{eq}_F(E_{p''})(1-f^{eq}_F(E_{p'}))(1-f^{eq}_F(E_{p'''}))\left(\chi_+(p_z')+\chi_-(p_z''')-\chi_+(p_z)-\chi_-(p_z'')\right),\nonumber\\\label{qbarq}
\eear
where the matrix element is given by a similar expression as (\ref{m}),
\be
{\cal M}=g_s^2\int{dq^1\over (2\pi)} G_{\mu\nu}(q) (R_{00}({\bm q}_\perp))^2 e^{-i{q^1\over 2eB}\left(p_2+p_2'+p_2''+p_2'''\right)}[\bar u(p_z')\gamma^\mu_\parallel u(p_z)]
[\bar v(p_z'')\gamma^\nu_\parallel v(p_z''')]\,,
\ee
where ${\bm q}={\bm p}'-{\bm p}={\bm p}''-{\bm p}'''$.
We neglected color indices, and the color trace simply gives a factor of $T_R C_2(R)$ in front of $|{\cal M}|^2$.
The (color stripped) full gluon propagator $G_{\mu\nu}(q)$ including the dominant 1-loop self energy from hard thermal LLL loop
can be found in Ref.\cite{Fukushima:2015wck,Hattori:2012je,Hattori-Satow,Li:2016bbh}.
This self energy is of order $\sim\alpha_s eB$, and dominates over the one from hard thermal gluons of $\alpha_s T^2$. 
Note that $(\mu,\nu)$ in the above are projected to 1+1 dimensions due to the coupling to LLL quarks, and in this case the re-summed gluon propagator is given by 
\be
G_{\mu\nu}(q)={i\eta^\parallel_{\mu\nu}\over -q^2+m_{D,B}^2\changed{+i{\text{Im}}\varPi_\parallel(q)}}={i\eta^\parallel_{\mu\nu}\over {\bm q}_\perp^2-q_\parallel^2+m_{D,B}^2\changed{+i{\text{Im}}\varPi_\parallel(q)}}\,,
\label{eq:Gprop}
\ee
where $\eta^{\mu\nu}_\parallel = {\rm diag} (1,0,0,-1)$, \changed{and $\text{Im}\varPi_\parallel(q)$ is the imaginary part of the gluon self-energy which we have retained for later convenience.}  
In Eq.~(\ref{eq:Gprop}), we have dropped the gauge-dependent terms, 
which are proportional to the momentum and vanish at on-shell kinematics due to the 1+1 dimensional Ward identity. 
The Debye mass from the \changed{real part of the self-energy from the LLL loop in small quark mass limit} is 
\be
m_{D,B}^2=4\alpha_s T_RN_F\left(eB\over 2\pi\right)e^{-{{\bm q}_\perp^2\over2eB}}\,.
\ee

\changed{We will see later that $-q_\parallel^2$ becomes small in the infrared region $q_z\ll T$ where we can get potentially infrared enhanced contributions.
In this region, the imaginary part of the self-energy of the exchanged gluon becomes large:
it is estimated as~\cite{Fukushima:2015wck} ${\text{Im}}\varPi_\parallel(q) \sim m^2_{D,B}T/q_z$.
This is much larger than $m^2_{D,B}$ when $q_z\ll T$, and becomes much larger than $-q^2_\parallel$ when $|q_z|\ll {\Lambda}_{\text{IR}}\equiv(m_{D,B}/m_q)^{2/3}T$, using the fact $-q_\parallel^2\sim m_q^2q_z^2/T^2$ (see Eq.(\ref{qparallel})).
Thus, the imaginary part of the gluon self-energy introduces a new dynamical IR cutoff, ${\Lambda}_{\text{IR}}$, for this infrared sensitive region, replacing the Debye mass $m_{D,B}$.
Because the purpose of the current calculation is just to make an order estimate, we simply take into account this effect by replacing the terms coming from the gluon self-energy in the exchanged gluon propagator as $m_{D,B}^2+i{\text{Im}}\varPi_\parallel(q)\rightarrow {\Lambda}^2_{\text{IR}}$.
}

For t-channel scatterings, the momentum exchange is space like $-q_\parallel^2>0$. 
Note that the LLL quarks inside the one-loop gluon self energy couple only to one of the two transverse modes of gluons which has the in-plane polarization 
with respect to the plane spanned by the external magnetic field and the gluon momentum. The other polarization mode perpendicular to the plane is decoupled from the LLL quarks, and is not affected by the LLL self-energy. This can be seen by the vanishing current transverse to $\bm B$, $\bar\psi\gamma_\perp\psi=0$, in the LLL approximation (see Appendix A).
Therefore, whereas the in-plane mode is screened by $ m_{D,B}$, 
the other mode, the out-of-plane polarization, is not screened by the LLL quark loop. 
Nevertheless, as long as LLL quarks are involved in the scatterings as carriers and/or scatterers, 
one can use Eq.~(\ref{eq:Gprop}) because the out-of-plane mode is decoupled from LLL quarks, and
will not appear anyway (Eq.~(\ref{eq:Gprop}) contains only in-plane mode contributions).
However, for the gluon-gluon scattering considered around Eq.~(\ref{eq:damping-gg}), 
one needs to include the ordinary gluon-loop contribution for out-of-plane modes,
which gives rise to the screening mass $m_T^2 \sim \alpha_s T^2$ for the out-of-plain polarization mode.

Note the $q^1$ integral in ${\cal M}$ and we need to compute $|{\cal M}|^2$: the Schwinger phase will play a crucial role to get the right LLL density of states as follows.
The $|{\cal M}|^2$ is
\be
|{\cal M}|^2=g_s^4\int {dq^1\over (2\pi)}\int{d\tilde q^1\over (2\pi)}
 e^{-i{(q^1-\tilde q^1)\over 2eB}\left(p_2+p_2'+p_2''+p_2'''\right)}\times ({\rm other\,\,stuff})
 \,.\ee
Consider $p_2$ integrals in (\ref{qbarq}) with the above Schwinger phase,
\be
\int {dp_2'\over (2\pi)}\int {dp_2''\over (2\pi)}\int {dp_2'''\over (2\pi)} (2\pi)\delta(p_2+p_2''-p_2'-p_2''')\,e^{-i{(q^1-\tilde q^1)\over 2eB}\left(p_2+p_2'+p_2''+p_2'''\right)}\,.
\ee
Without loss of generality, we can choose $p_2=0$, and the ``other stuff'' depends only on $q_2=p_2'=p_2''-p_2'''$, so
after performing $p_2'''$ integral using the $\delta$-function, we have
\be
\int {dq_2\over (2\pi)}\int {dp_2''\over (2\pi)}e^{-i{(q^1-\tilde q^1)\over eB}\cdot p_2''}=\left(eB\over 2\pi\right)(2\pi)\delta(q^1-\tilde q^1)\int {dq_2\over (2\pi)}\,.
\ee
Then, we have
\bear
&&\int {dp_2'\over (2\pi)}\int {dp_2''\over (2\pi)}\int {dp_2'''\over (2\pi)} (2\pi)\delta(p_2+p_2''-p_2'-p_2''')\,e^{-i{(q^1-\tilde q^1)\over 2eB}\left(p_2+p_2'+p_2''+p_2'''\right)}|{\cal M}|^2\nonumber
\\ &=& g_s^4 \left(eB\over 2\pi\right) \int {d^2{\bm q}_\perp\over (2\pi)^2} \times ({\rm other\,stuff})\,,
\eear
where the other stuff depends on ${\bm q}_\perp=(q_1,q_2)$ only. This way, the correct LLL density of states $(eB/2\pi)$ is produced.

As explained in the previous subsection, the 1+1 dimensional energy-momentum condition fixes the unique solution
$p_z'=p_z''$ and $p_z'''=p_z$, and working out Jacobian, we have
\bear
&&\int{dp_z'\over (2\pi)}\int{dp_z''\over (2\pi)}\int{dp_z'''\over (2\pi)} (2\pi)\delta(p_z+p_z''-p_z'-p_z''')(2\pi)\delta(E_p+E_{p''}-E_{p'}-E_{p'''})\nonumber\\&=&\int{dp_z'\over (2\pi)}{E_p E_{p'}\over |E_pp_z'-E_{p'}p_z|}\,.\label{jac}
\eear
Also, the t-channel momentum exchange becomes
\be
-q_\parallel^2=-(p'-p)^2_\parallel=2(E_p E_{p'}-p_zp_z'-m_q^2)>0\,,
\ee
and finally the spinor trace is easily computed to be
\be
\Big([\bar u(p_z')\gamma^\mu_\parallel u(p_z)][\bar v(p_z')\gamma_{\parallel  \mu} v(p_z)]\Big)^2=16 m_q^4\,.
\ee

We have all the ingredients to write down the collision term as
\bear
C[f_+(p_z)]&=&2 g_s^4T_RC_2(R)\left(eB\over 2\pi\right)(16 m_q^4){1\over (2E_p)^2}\beta f^{eq}_F(E_p)(1- f^{eq}_F(E_p))
\nonumber\\&\times&\int{dp_z'\over (2\pi)} {1\over (2E_{p'})^2}{E_pE_{p'}\over |E_pp_z'-E_{p'}p_z|}\int{d^2{\bm q}_\perp\over(2\pi)^2} e^{-{{\bm q}_\perp^2\over eB}}{1\over \left({\bm q}_\perp^2+(-q_\parallel^2+\changed{{\Lambda}^2_{\text{IR}}})\right)^2} \nonumber\\&\times& f^{eq}_F(E_{p'})(1- f^{eq}_F(E_{p'})) (\chi_+(p_z')-\chi(p_z))\nonumber\\&=&
8\pi\alpha_s^2T_RC_2(R)\left(eB\over 2\pi\right)m_q^4{\beta\over E_p} f^{eq}_F(E_p)(1- f^{eq}_F(E_p))
\nonumber\\&\times&\int{dp_z'\over (2\pi)}{1\over E_{p'}}{1\over |E_pp_z'-E_{p'}p_z|}{1\over (\changed{{\Lambda}^2_{\text{IR}}}+2(E_pE_{p'}-p_zp_z'-m_q^2))}\nonumber\\&\times& f^{eq}_F(E_{p'})(1- f^{eq}_F(E_{p'})) (\chi_+(p_z')-\chi(p_z))\,,\label{col22}
\eear
where we neglected $e^{-{{\bm q}_\perp^2\over eB}}$ since ${\bm q}_\perp^2\sim \changed{{\Lambda}^2_{\text{IR}}}-q_\parallel^2\lesssim T^2\ll eB$.
Although the expression  looks proportional to $m_q^4$, they cancel in a subtle way in the final result.
We need to consider two separate cases; 1) $p_z\cdot p_z'<0$ (chirality flip) and 2) $p_z\cdot p_z'>0$ (chirality non-flip).

1) $p_z\cdot p_z'<0$ case (chirality flip): In this case, $|E_pp_z'-E_{p'}p_z|$ and $-q_\parallel^2$ are all of order $p_z\sim T$, and there is no IR enhanced regime in the integration. Typical momenta of integration is of order $T$ and therefore, the collision term in this case is of order
\be
C\sim \alpha_s^2 (eB){m_q^4\over T^6}\,\, \chi_+\,,
\ee
which is clearly sub-leading to the 1-2 process, $C_{1-2}\sim \alpha_s m_q^2/T^2 \chi_+$, when $\alpha_seB\ll (m_q^2,T^2)$.

2) $p_z\cdot p_z'>0$ case (chirality non-flip): In this case, both $|E_pp_z'-E_{p'}p_z|$ and $-q_\parallel^2$ are proportional to $m_q^2$, canceling the $m_q^4$ factor in the numerator. Explicitly, it is better to write them as
\be
 |E_pp_z'-E_{p'}p_z|={m_q^2|p_z'^2-p_z^2|\over  E_pp_z'+E_{p'}p_z}\,,
\ee
and 
\be
-q_\parallel^2={2m_q^2 (p_z'-p_z)^2\over E_pE_{p'}+p_zp_z'+m_q^2}\,,\label{qparallel}
\ee
that shows the $m_q^2$ factor clearly.
In (\ref{col22}), when we compare $\changed{{\Lambda}^2_{\text{IR}}}$ and $-q_\parallel^2$, the $-q_\parallel^2$ dominates for most of the $p_z'$
since $\changed{{\Lambda}^2_{\text{IR}}}\ll m_q^2$, unless $p_z'$ is very close to $p_z$ (IR region) so that
\be
|p_z'-p_z|^2\lesssim {\changed{{\Lambda}^2_{\text{IR}}}\over m_q^2} T^2\ll T^2\,,
\ee
below which $\changed{{\Lambda}^2_{\text{IR}}}$ dominates.
We will find a logarithmic enhancement from the range \changed{$\Lambda_{IR}T/m_q\ll |p_z'-p_z|\ll T$}.
To identify this logarithm near this IR region, it is safe to approximate $p_z'\approx p_z$ in the integrand except IR diverging pieces, and
we also have a ``diffusion'' expansion 
\be
\chi_+(p_z')-\chi_+(p_z)=(p_z'-p_z)\partial_{p_z}\chi_+(p_z)+{1\over 2}(p_z'-p_z)^2\partial_{p_z}^2 \chi_+(p_z)\,.
\ee
Putting all these in (\ref{col22}), we finally have
\bear
C[f_+(p_z)]_{\rm LLog}&\approx& (4\pi)\alpha_s^2 T_RC_2(R)\left(eB\over 2\pi\right){1\over T}\int_{|p_z'-p_z|={\Lambda_{IR}\over m_q}T}^T {dp_z'\over (2\pi)}{1\over |p_z'-p_z|}\nonumber\\ &\times&\partial_{p_z}\left(E_p(f^{eq}_F(E_p))^2\left(1-f^{eq}_F(E_p)\right)^2 \partial_{p_z}\chi_+(p_z)\right)
\nonumber\\&=&
\changed{2\over 3}\alpha_s^2 T_RC_2(R)\left(eB\over 2\pi\right){1\over T}\log\left(\changed{m_q^5\over \alpha_s eB T^3}\right) \nonumber\\&\times& \partial_{p_z}\left(E_p(f^{eq}_F(E_p))^2\left(1-f^{eq}_F(E_p)\right)^2 \partial_{p_z}\chi_+(p_z)\right)\,,\nonumber\\\eear
which is what we claimed in (\ref{2to2}). It is easy to check that the other region $|p_z'-p_z|\lesssim\changed{ {\Lambda_{IR}\over m_q}T}$ (where $\changed{{\Lambda}^2_{\text{IR}}}$ dominates over $-q_\parallel^2$)
produces only a constant under the above log.

Finally let us see what happens when $m_q^2\ll \alpha_s eB$. 
\changed{In this case, we see that ${\Lambda}^2_{\text{IR}}$ always dominates over $-q_\parallel^2$ in (\ref{col22}), and there is no infrared enhancement in the $q$ integral, which means that $q$ integral is dominated by the region $q\sim T$. Then, the real and imaginary part of the gluon self-energy is of the same order to be $\sim m_{D,B}^2\sim\alpha_s eB$. We therefore have the replacement ${\Lambda}^2_{\text{IR}}\to \alpha_s eB$} in this case, and have the estimate
 (for chirality non-flip case $p_z\cdot p_z'>0$)
\be
C\sim \alpha_s^2 (eB/T^2)m_q^2 {1\over \alpha_s eB}\int dp_z' {1\over |p_z'-p_z|}(\chi_+(p_z')-\chi(p_z))\,.
\ee
With the expansion of $(\chi_+(p_z')-\chi(p_z))\approx (p_z'-p_z)\partial_{p_z}\chi_+(p_z)$ for small $(p_z'-p_z)$, there is no IR enhancement in the above expression, and we have the estimate \be
C\sim \alpha_s (m_q^2/T^2)\chi_+\,,
\ee
which is of the same order as the 1-to-2 process (\ref{1to2app}). Therefore, this has to be included as well in leading order computation, as we claimed in the introduction.

\subsection*{Quark-gluon t-channel scattering}

Another 2-to-2 process of interest is quark-gluon t-channel scattering with a gluon exchange.
One can check by an explicit computation that the t-channel fermion exchange is sub-leading to this process by a factor of $\log(T^2/\alpha_seB)$, and we skip its detail here.
We consider quark-gluon scattering: ${\bm p}+{\bm k}\to {\bm p'}+{\bm k'}$ with the momentum exchange ${\bm q}={\bm p'}-{\bm p}={\bm k}-{\bm k'}$. The collision term from this is
\bear
C[f_+(p_z)]&=& {1\over 2E_p}\int_{\bm p'}\int_{\bm k}\int_{\bm k'} |{\cal M}|^2 (2\pi)^2\delta^{(2)}({\bm p}+{\bm k}-{\bm p'}-{\bm k'})(2\pi)\delta(E_p+E_k-E_{p'}-E_{k'})\nonumber\\&\times&
\beta f^{eq}_F(E_p)f^{eq}_B(E_k)(1-f^{eq}_F(E_{p'}))(1+f_B^{eq}(E_{k'}))(\chi_+(p_z')-\chi_+(p_z))\,,\label{2-2glu}
\eear
where we dropped the gluon distribution $\chi_g(k)=0$, and $E_k=|{\bm k}|$ is the gluon on-shell energy.
The matrix element is
\bear
{\cal M}&=&g_s^2 f^{abc}[\bar u(p_z')t^a \gamma^\beta_\parallel u(p_z)]G_{\beta\mu}(q)\left((k-2q)_\nu \eta_{\mu\alpha}
+(k_\alpha+q_\alpha)\eta_{\mu\nu}-(2k-q)_\mu\eta_{\alpha\nu}\right)\nonumber\\&\times&
\epsilon^\nu(\tilde\epsilon^\alpha)^* R_{00}({\bm q}_\perp) e^{i\Sigma}\,,
\eear
where $(\epsilon,\tilde\epsilon)$ are polarizations of incoming and outgoing gluons, $a,b,c$ are color indices of exchanged, incoming and outgoing gluons respectively.
Since the exchanged gluon propagator $G_{\beta\mu}(q)$ is coupled to the purely longitudinal quark current on one side, one can show that the only non-vanishing part is
\be
G_{\beta\mu}(q)={i\eta^\parallel_{\beta\mu}\over {\bm q}_\perp^2-q_\parallel^2+m_{D,B}^2
\changed{+i{\text{Im}}\varPi_\parallel(q)}}\approx {i\eta^\parallel_{\beta\mu}\over {\bm q}_\perp^2-q_\parallel^2+\changed{{\Lambda}^2_{\text{IR}}}}\,,
\ee
as before.
Color traces in $|{\cal M}|^2$ produces a color factor of $N_c C_2(R)$, and other than this we can forget about colors in the following. 

We will show that the above collision integral has a logarithmic IR enhancement from $m_{D,B}\ll q\ll T$,
and for this leading log order, we can focus on only leading terms in small $q/T\ll 1$ expansion in the matrix element.
For example, in leading order of $q/T$ we have
\be
\left((k-2q)_\nu \eta_{\mu\alpha}
+(k_\alpha+q_\alpha)\eta_{\mu\nu}-(2k-q)_\mu\eta_{\alpha\nu}\right)
\epsilon^\nu(\tilde\epsilon^\alpha)^*\approx -2 k^\parallel_\mu (\epsilon\cdot\tilde\epsilon^*)+{\cal O}(q)\,
\ee
using the fact that $\epsilon\cdot k=\tilde\epsilon\cdot k=0+{\cal O}(q)$.
Moreover, since $k\approx k'+{\cal O}(q)\sim T$, the polarization summation in $|{\cal M}|^2$ gives
\be
\sum_{\epsilon,\tilde\epsilon} |(\epsilon\cdot\tilde\epsilon^*)|^2=2+{\cal O}(q/T)\,.
\ee
Performing spinor trace, we have (we can neglect $R_{00}({\bm q}_\perp)$ since ${\bm q}_\perp^2 \ll T^2\ll eB$)
\bear
|{\cal M}|^2&=& 16g_s^4 N_c C_2(R){1\over |{\bm q}_\perp^2-q_\parallel^2+\changed{{\Lambda}^2_{\text{IR}}}|^2}\left(2(k\cdot p)(k\cdot p')
-(p\cdot p'-m_q^2)k^2_\parallel\right)\,.
\eear
Also, a short algebra gives
\be
p\cdot p'-m_q^2={m_q^2 q_z^2\over E_pE_{p'}+p_z p_z'+m_q^2}\sim {\cal O}(q^2)\,,
\ee
since for small momentum transfer $q_z\ll p_z\sim T$, we have $p_z\cdot p_z'>0$. Therefore, we can neglect $(p\cdot p'-m_q^2)$ in the above as well, and we have
\be
|{\cal M}|^2\approx 32 g_s^4 N_c C_2(R){(E_kE_p-k_z p_z)^2\over |{\bm q}_\perp^2-q_\parallel^2+\changed{{\Lambda}^2_{\text{IR}}}|^2}\,.
\ee
The phase space integral in (\ref{2-2glu}) is also approximated in small $q/T$ limit as follows.
We first re-parametrize ${\bm k'}$ integral to the momentum transfer ${\bm q}\equiv {\bm k}-{\bm k'}$,
\be
\int_{\bm k'}=\int{d^3{\bm q}\over (2\pi)^3 2E_{k-q}}\,,
\ee
and performing ${\bm p'}$ integral to arrive at
\bear
C[f_+(p_z)]&=& {1\over 2E_p}{1\over 2E_{p+q}}\int {d^3{\bm k}\over (2\pi)^3 2E_k}\int {d^3 {\bm q}\over (2\pi)^3 2E_{k-q}}
|{\cal M}|^2 (2\pi)\delta(E_p+E_k-E_{p+q}-E_{k-q})\nonumber\\
&\times& \beta f^{eq}_F(E_p)f^{eq}_B(E_k)(1-f^{eq}_F(E_{p+q}))(1+f^{eq}_B(E_{k-q}))(\chi_+(p_z+q_z)-\chi_+(p_z))\,.\nonumber\\
\eear 
Taking the small $q/T\ll 1$ limit, we have
\be
\delta(E_p+E_k-E_{p+q}-E_{k-q})=\delta\left(\hat{\bm k}\cdot{\bm q}-{p_z\over E_p}q_z\right)\,,
\ee
and 
\be -q^2_\parallel=-(q^0)^2+q_z^2=\left(1-{p_z^2\over E_p^2}\right)q_z^2={m_q^2 \over E_p^2} q_z^2\,.
\ee

By neglecting subleading terms in $q/T$, we have
\bear
C[f_+(p_z)]&=& {1\over (2E_p)^2}\int {d^3{\bm q}\over (2\pi)^3}\int {d^3{\bm k}\over (2\pi)^3 (2E_k)^2} |{\cal M}|^2
(2\pi)\delta\left(\hat{\bm k}\cdot{\bm q}-{p_z\over E_p}q_z\right)\label{semifin}\\&\times&\beta
f^{eq}_F(E_p)f^{eq}_B(E_k)(1-f^{eq}_F(E_{p}))(1+f^{eq}_B(E_{k}))(\chi_+(p_z+q_z)-\chi_+(p_z))\,,\nonumber
\eear
where
\be 
|{\cal M}|^2=32 g_s^4 N_c C_2(R) {(E_kE_p-k_zp_z)^2\over ({\bm q}_\perp^2+{m_q^2\over E_p^2}q_z^2+\changed{{\Lambda}^2_{\text{IR}}})^2}\,.\label{apf1}
\ee
The phase space $\bm k$ integral is doable with some effort. The result is
\bear
&&\int {d^3 {\bm k}\over (2\pi)^3 (2E_k)^2} (2\pi)\delta\left(\hat{\bm k}\cdot{\bm q}-{p_z\over E_p}q_z\right)(E_kE_p-k_zp_z)^2
\beta f^{eq}_B(E_k)(1+f^{eq}_B(E_k))\nonumber\\&=&
{\pi T^2\over 48}{1\over |{\bm q}|}{1\over E_p^2}\left(2m_q^4+5 m_q^2 p_z^2\left(1-{q_z^2\over |\bm q|^2}\right)+3p_z^4\left(1-{q_z^2\over |\bm q|^2}\right)^2\right)\,.\nonumber\\\label{apf2}
\eear
With this, one easily recognizes that there is an IR divergence in (\ref{semifin}) in small ${\bm q}$ regime that is regulated by $\changed{{\Lambda}^2_{\text{IR}}}$. To extract this leading log enhancement, 
we first do the ``diffusion expansion'' for small $q_z\ll p_z$:  $\chi_+(p_z+q_z)-\chi_+(p_z)\approx q_z \partial_{p_z}\chi_+(p_z)+{1\over 2}q_z^2 \partial_{p_z}^2\chi_+(p_z)$, and only the quadratic piece contributes due to symmetry $q_z\to-q_z$ in the integrand (however, see the discussion at the end of this subsection). We also rescale $q_z\to {E_p\over m_q}q_z$ to finally obtain
\bear
C[f_+(p_z)]&=& {\pi\over 3}\alpha_s^2 T^2 N_c C_2(R) f^{eq}_F(E_p)(1-f_F^{eq}(E_p)) S\left({m_q/ E_p}\right) \left(\int_0^{\sim T} d|{\bm q}|{|{\bm q}|^3\over (|{\bm q}|^2+\changed{{\Lambda}^2_{\text{IR}}})^2}\right)\partial_{p_z}^2\chi_+(p_z)\nonumber\\
&=&{\pi\over 9}\alpha_s^2 T^2 N_c C_2(R) f^{eq}_F(E_p)(1-f_F^{eq}(E_p)) S\left({m_q/ E_p}\right)\changed{\log\left(m^2_q\over \alpha_s eB\right) }
\partial_{p_z}^2\chi_+(p_z)\,,\label{almostfin}
\eear
where the UV cutoff is provided by $T$ \changed{above which} our low $q/T\ll 1$ approximation is no longer valid, and the dimensionless function $S(m_q/E_p)$ is defined by the angular integral,
\bear
S(x)&=&x^2 \int_{-1}^{+1}d\cos\theta{\cos^2\theta\over\sqrt{x^2\sin^2\theta+\cos^2\theta}}\\&\times&\left(2+5(1-x^2){\sin^2\theta\over (x^2\sin^2\theta+\cos^2\theta)}+3(1-x^2)^2{\sin^4\theta\over (x^2\sin^2\theta+\cos^2\theta)^2}\right)\,.\nonumber
\eear
In the small mass limit of $x=m_q/E_p\to 0$, the integral is dominated by the region $\cos\theta\sim x$ (and $\sin\theta\sim 1$), and parametrizing this region as $\cos\theta=xt$, we have
\be
S(x)\to \int_{-\infty}^{+\infty} dt {3t^2\over (1+t^2)^{5\over 2}}=2\,,\quad x\to 0\,.
\ee

Although we derived (\ref{almostfin}) by neglecting $q/T$ corrections in the matrix element, there can be other contributions
of the same order that result from a first order $q/T$ correction from the matrix element combined with the linear term of the ``diffusion expansion''
of $\chi_+(p_z+q_z)-\chi_+(p_z)$. The easiest way to handle this subtlety is to work with the variational functional $I$ such that $C={\delta I\over\delta \chi_+}$ as in literature \cite{Arnold:2000dr}. The upshot of this exercise is that the final leading log collision term is a total derivative,
\bear
C[f_+(p_z)]&=&{\pi\over 9}\alpha_s^2 T^2 N_c C_2(R)
\changed{\log\left(m^2_q\over \alpha_s eB\right) }
\partial_{p_z}\Big( f^{eq}_F(E_p)(1-f_F^{eq}(E_p)) S\left({m_q/ E_p}\right)\partial_{p_z}\chi_+(p_z)\Big)\,,\nonumber\\\label{refin}
\eear
which is our final result in this section. As claimed before, this collision term is parametrically smaller than the 1-to-2 process
when $m_q^2\gg \alpha_s eB\gg \alpha_s T^2$. 

\vfil

\end{document}